\documentclass[3p,times,pdflatex]{elsarticle}
\input{header}
\makeatletter
\def\ps@pprintTitle{%
  \let\@oddhead\@empty
  \let\@evenhead\@empty
  \def\@oddfoot{\reset@font\hfil\thepage\hfil}
  \let\@evenfoot\@oddfoot
}
\makeatother
\makeatletter
\renewcommand{\MaketitleBox}{%
  \resetTitleCounters
  \def\baselinestretch{1}%
  \begin{center}
    \def\baselinestretch{1}%
    \Large \@title \par
    \vskip 18pt
    \normalsize\elsauthors \par
    \vskip 10pt
    \footnotesize \itshape \elsaddress \par
  \end{center}
  \vskip 12pt
}
\makeatother
\newcommand\snowmass{\begin{center}\rule[-0.2in]{\hsize}{0.01in}\\\rule{\hsize}{0.01in}\\
\vskip 0.1in Submitted to the  Proceedings of the US Community Study\\ 
on the Future of Particle Physics (Snowmass 2021)\\ 
\rule{\hsize}{0.01in}\\\rule[+0.2in]{\hsize}{0.01in} \end{center}}

\usepackage{etoolbox}
\patchcmd{\MaketitleBox}{\footnotesize\itshape\elsaddress\par\vskip36pt}{\footnotesize\itshape\elsaddress\par\parbox[b][36pt]{\linewidth}{\vfill\hfill\textnormal{\today}\hfill\null\vfill}}{}{}%
\patchcmd{\pprintMaketitle}{\footnotesize\itshape\elsaddress\par\vskip36pt}{\footnotesize\itshape\elsaddress\par\parbox[b][36pt]{\linewidth}{\vfill\hfill\textnormal{\today}\hfill\null\vfill}}{}{}%
\begin{document}
\begin{frontmatter}
\title{\bf Snowmass 2021 Instrumentation Frontier (IF5 - MPGDs) - White Paper 2: \\  Micro Pattern Gaseous Detectors for Nuclear Physics. }
\author[7]{F. Barbosa}
\author[1]{D. Bazin}
\author[4]{F. Boss\`u}
\author[1]{M. Cortesi}
\author[3]{S. Dalla Torre}
\author[7]{S. Furletov} 
\author[7]{Y. Furletova} 
\author[1]{P. Gueye}
\author[7]{K. Gnanvo (Coordinator)\corref{cor1}}
\ead{kagnanvo@jlab.org}
\author[2]{M. Hohlmann}
\author[1]{W. Mittig}
\author[4]{D. Neyret}
\author[6]{M. Posik (Coordinator)\corref{cor1}}
\ead{posik@temple.edu}
\author[5]{C. Wrede}
\address[1]{Facility for Rare Isotope Beams, Michigan State University, East Lansing, MI 48824, USA}
\address[2]{Florida Institute of Technology, Melbourne, FL 32901, USA}
\address[3]{INFN Trieste, Via A. Valerio, 2
34127 Trieste, Italy}
\address[4]{IRFU, CEA, Universit\'e Paris-Saclay, F-91191 Gif-sur-Yvette, France}
\address[5]{Department of Physics and Astronomy, Michigan State University, East Lansing, MI 48824, USA}
\address[6]{Temple University, Philadelphia, PA 19122, USA}
\address[7]{Thomas Jefferson National Accelerator Facility, Newport News, VA 23606, USA}

\today
%
\end{frontmatter}
\snowmass
\tableofcontents
\renewcommand{\baselinestretch}{1.02}
\newcommand{\plot}[5]{
 \begin{figure}[#1]
 \begin{center}
 \includegraphics[width=#2]{#3}
 \end{center}
 \caption{#4}\label{#5}
 \end{figure}}
 \newcommand{\plots}[6]{
 \begin{figure}[#1]
 \begin{center}
 \begin{tabular}{cc}
 \includegraphics[width=#2]{#3} & 
 \includegraphics[width=#2]{#4}
 \end{tabular}
 \end{center}
 \caption{#5}\label{#6}
 \end{figure}}

\newpage
\section{Executive Summary}
\vspace{-3mm}
Many current and future nuclear physics (NP) experiments across the United States have and are implementing micro-pattern gas detectors (MPGDs) to be used for tracking and PID purposes. MPGDs are capable of operating in high rate environments and providing excellent spatial resolution over a large-area with a low material budget. Summarized in this white paper is the role that MPGDs are playing in NP experiments and the R\&D which is needed to meet the requirements of future NP experiments.   
\vspace{-3mm}
\subsection{{\bf Advanced Micro Pattern Gas Detectors for Tracking at the Electron-Ion Collider}}
\vspace{-2.5mm}
Successfully completing the physics program of the Electron-Ion Collider (EIC), to be built at Brookhaven National Laboratory, requires its tracking system to have low mass ($ X/X_0 \lesssim 1\% $), large area $\mathcal{O}$($1 m$), and excellent spacial resolution $\mathcal{O}$($100 \mu m$). MPGDs, such as GEM, Micromegas, and $\mu$RWELL can meet these requirements. Furthermore, the EIC is expected to have relatively low rates, below 100 $kHz/cm^2$, which is well within the operating range of current MPGDs. Current R\&D is focused on reducing large-area detector material budgets and reducing the number of channels needed to be read out while maintaining excellent spatial resolution. The EIC can benefit from future R\&D which looks to further reduce the material and service budget of the detector, pushing the spatial resolution down towards the order of $\sim 20 \mu m$, and implementing particle identification capabilities into MPGD detectors. 
\vspace{-3mm}
\subsection{\bf{MPGD Technologies for Low Energy Nuclear Physics at FRIB}}
\vspace{-2.5mm}
The Facility for Rare Isotope Beams (FRIB) at Michigan State University (MSU) will become the world’s most advanced facility for the production of rare isotope beams (RIBs). With the delivery of beams starting in Spring 2022, FRIB will be capable of producing a majority (around 80\%) of the isotopes predicted to exist including more than 3,000 new isotopes, opening exciting perspectives for exploring the uncharted regions of the nuclear landscape. Its scientific impact will span a better understanding of open quantum systems at the limits of stability through investigations of the structure and reactions of atomic nuclei and their roles in nuclear astrophysics, low-energy tests of fundamental symmetries, and practical applications that benefit humanity. MPGD technologies play an important role for the success of the science program at FRIB. Applications of MPGD technologies include low-pressure tracking and particle-identification (PID) at the focal planes of magnetic spectrometers, Active-Target Time-Projection-Chambers (TPCs), and TPCs for the detection of exotic decay modes with stopped RIBs. The unprecedented discovery potential of FRIB can be achieved by implementing state-of-the-art experimental equipment and overcoming challenges of present devices by taking the following measures: improving spatial and energy resolutions, optimizing pure-gas operation for active target mode, improving reliability and radiation hardness at a lower cost, reducing ion-back flow to minimize secondary effects and increase counting rate capability, and integrating electronic readout to reach high channel density, fast data processing and storage.
\vspace{-3mm}
\subsection{\bf{MPGD Technologies for Nuclear Physics at Jefferson Lab}}
\vspace{-2.5mm}
Future spectrometer for NP experiments at the Thomas Jefferson National Accelerator Facility (Jefferson Lab) require large area $\mathcal{O}$(m),  low mass (X/X$_0 \leq$ 1\%), excellent spatial  $\mathcal{O}(100\, \muup \rm{m})$, excellent timing $\mathcal{O}$(10 ns), high rate $\mathcal{O}$(1MHz / cm$^2$) tracking detectors technologies for operation in high background rate and high radiation environment. Only MPGD technologies such as Gas Electron Multipliers (GEMs), Micro Mesh Gaseous (Micromegas) or Resistive Micro Well ($\mu$RWELL) detectors are able to satisfy the challenges of high performances for large acceptance at reasonably low cost. Critical R\&D for the next decades will focus on new ideas to develop ultra-low mass, large area and radiation tolerant MPGD trackers operating at high rate capabilities. Performance of new  materials (Chromium GEMs, Aluminum based readout strips) will be investigated as well as original concepts for anode readout such as capacitive and resistive and zigzag readouts for high-performance \& low channel count MPGD detectors.  
\vspace{-3mm}
\subsection{\bf{MPGD Technologies for Particle Identification in Nuclear Physics Experiments}}
\vspace{-2.5mm}
%
PID plays a very important role for high energy physics (HEP) and NP physics. The next generation of high intensity accelerators and high demand for precision measurements from the physics will require the development of high granularity detectors, such as those based on MPGD technologies. Combining a high precision tracker with PID capable technology could prove valuable for future experiments. 
A high precision MPGD tracker combined with a transition radiation (TR) option for particle identification could provide important information necessary for electron identification and hadron suppression. A radiator, installed in front of a MPGD entrance window provides an efficient yield of TR photons. 
MPGD-based photon detectors offer the ability to provide PID through Cerenkov imaging techniques. Such detectors are attractive for experiments like the EIC as they can offer a cost efficient option for large-area, low material budget detector with the ability to operate in a magnetic field.
%
%
\vspace{-3mm}
\subsection{\bf{Electronics, DAQ, and 
Readout Systems for MPGD Technologies}}
\vspace{-2.5mm}
The EIC will implement a full streaming readout architecture. This trigger-less implementation will consist of front-end circuitry and processors to enable data collection, processing and analysis: front-end ASICs will be designed to meet wide bandwidth sub-detector and system requirements; front-end processors will include FPGAs to provide data aggregation and enable flexible algorithms to reduce data volume while maintaining wide system bandwidth; system clock distribution with timing precision of the order of 1 ps; link exchange modules and servers for data processing; and data transport via extensive use of optical fibers. The development of ML/AI algorithms will play a critical role in enabling a full detector bandwidth of 100 Tbps and to deliver data output rates of 100 Gbps.
Readout of MPGD detectors requires specific front-end ASICs able to amplify and digitize the detector signals with performance requirements depending of its constraints and application. The ASICs should be also compatible with the high-speed streaming readout DAQ systems that are considered for the future experiments at EIC and elsewhere. Present chips, like SAMPA or VMM, partially satisfy these requirements and can be used for specific applications. Nevertheless, an initiative is launched to develop a new versatile ASIC covering most of the constraints of the different MPGD applications in HEP experiments. The new SALSA chip is meant to be equally adapted to the streaming and triggered readout paradigms and is being designed in a more modern 65 nm technology. In addition, the SALSA chip holds a promise to be the base of a new family of integrated circuits covering several specific applications requiring on-demand adaptation of some of its functionalities.
\vspace{-3mm}
\subsection{\bf{Need for Dedicated Nuclear Physics MPGD Development Facility}}
\vspace{-2.5mm}
As highlighted throughout this white paper, the detector requirements for NP experiments differ from those in HEP, which leads to  different detector R\&D paths and priorities. A dedicated US-based facility,  similar to the Gaseous Detector Development (GDD) lab at CERN or the SiDet facility at FNAL for silicon detector development, is strongly needed for MPGD groups of nuclear physics community. In addition, such a facility would benefit not only the NP community, but also for the broader particle physics community in the US.We envision such a facility to be hosted by one of DOE's National Laboratories, such as Jefferson Lab or Brookhaven National Laboratory.      
   
\newpage
\section{Advanced Micro Pattern Gas Detectors for Tracking at the Electron Ion Collider}

\subsection{The Electron-Ion Collider}
\subsubsection{Accelerator and Physics Overview} 
The Electron-Ion Collider (EIC) is a new facility being built at Brookhaven National Laboratory (BNL) in the U.S~\cite{CDR2021,EICYR2021}. The EIC is quite unique compared to other colliders, such as the LHC and RHIC. It will collide polarized electrons and polarized protons/ions at high luminosity ($10^{33}-10^{34}\; cm^{-2}s^{-1}$) and cover a wide range in center of mass energies ($\sqrt{s_{ep}}= 28 - 140\; GeV$). The beam energies will be asymmetric resulting in boosted kinematics, with higher activity occurring at larger pseudorapidity ($\eta > 1$, the hadron going direction). The bunch spacing will be $\sim 9\;ns$ and the beams will collide with a 25 mrad crossing angle. Another distinguishing feature of the EIC relative to other colliders is that it will have low particle multiplicity ($\lesssim$ 10 tracks/event), with an interaction rate of 500 kHz and insignificant pileup. Finally, the radiation environment will be much lower compared to the LHC (factor of $\sim100$).   
Taking advantage of the high beam polarizations, large luminosity, and wide kinematic coverage, the EIC will make high precision measurements over a broad range of physics topics~\cite{EICYR2021} via deep inelastic scattering (DIS), semi-inclusive DIS (SIDIS), and exclusive processes. To achieve this, an EIC detector must have a large pseudorapidity coverage ($|\eta|< 4$), and far forward subdetectors capable of going beyond $\eta = 4$, $4\pi$ coverage, a high-precision and low-mass tracking system, and excellent particle identification (PID) performance to separate $e,\;\pi,\;K$, and $p$ at the track level.   
\subsubsection{MPGD Tracking Requirements at the EIC} 
The unique environment of the EIC sets the tracking detector requirements, which differ significantly from other collider experiments. To provide the momentum resolution required by the physics program, the material budget available for tracking detectors is rather low, specifically below $1\%\  X/X_0$ per layer. This is needed to minimize multiple scattering in order to preserve electron and photon measurements. Finally, large-area tracking detectors can be used to ensure that the full kinematic acceptance is covered. Micropattern Gas Detectors (MPGDs), e.g.\ GEM, Micromegas, and $\muup$RWELL, are well suited for an EIC tracking system as they can be designed to meet these EIC tracking requirements. Furthermore, the EIC is expected to have particle rates well below $100\;kHz/cm^2$, which current MPGDs are capable of handling. 
The following subsections will summarize the current R\&D being carried out to produce large-area and low-mass MPGD trackers (Sec.\ref{sec:low_mass}), high resolution readout structures suited for the EIC (Sec.~\ref{sec:readout}), and future R\&D (Sec.~\ref{sec:future}). 
\subsection{Recent and Current R\&D Efforts}
In January 2011, Brookhaven National Laboratory (BNL), in association with Jefferson Lab and the DOE Office of Nuclear Physics established a generic detector R\&D program~\cite{eRD} to address the requirements for measurements at the EIC. From this program, the eRD3 and eRD6 projects were formed to focus on EIC compatible MPGD detectors, whose efforts have contributed greatly towards realizing high-resolution, large-area, and low-mass tracking detectors~\cite{9115073,Vandenbroucke2018,azmoun:2020tns,9638499, Hohlmann:2017sqj,Zhang:2017dqw,Zhang:2016vbk,Zhang:2015kxy,Zhang:2015pqa,Gnanvo:2014hpa,Gnanvo:2015xda}.
Current experiments such as Jefferson Lab's Super Bigbite Spectrometer (SBS) and CLAS12 experiments have already benefitted from this R\&D work. With the EIC moving towards becoming a reality after clearing CD0 and CD1 DOE milestones, a more targeted EIC R\&D program has been established with the eRD108 project focusing on further developing MPGD based detectors for the EIC experiment~\cite{eRD108Prop}. These R\&D activities have been focusing on two main aspects: reducing the material budget of large-area detectors and developing high-resolution readout structures with low-channel-counts.
\subsubsection{Accommodating a Small Material Budget}\label{sec:low_mass} 
To meet the momentum resolution and kinematic coverage requirements needed to successfully complete the EIC physics program, it is critical to reduce as much as possible the material of the tracking system, not only in the active area of the detectors, but also in its support structures. Passive structures are typically located within the tracking volume and can contribute to multiple scattering and acceptance holes. Building large-area detector modules or tiles reduces the amount of support material located in the tracking volume. However, building large-area detector modules while satisfying a low material budget is a non-trivial task since when detectors become larger they typically require more support. MPGD detectors, such as triple-GEMs are typically supported with G10/FR4 support material. A current R\&D task of eRD108 is investigating new materials which have lower radiation length and offer stronger tensile support.  

One way to reduce material in the active area of an MPGD tracker is to keep the standard amplification foils, such as GEM or $\mu$RWELL foils, but design the cathode and readout elements as Cu/Kapton based foils. Following this procedure, eRD6 was able to successfully build triple-GEM detectors with a length of roughly $1 m$ and radiation length of \mbox{$\sim 0.4\%\;X/X_0$} and operate and test them in beam, as shown in Fig.~\ref{fig:eicProto}. 
Applying this approach to building a $\mu$RWELL detector will further reduce the detector material budget below \mbox{$0.4\%\;X/X_0$} since the material of the amplification element of a $\mu$RWELL detector is less than that of a triple-GEM, and one can use the same foil-based cathode and readout elements as in the triple-GEM detector.

A significant contribution to the material budget of MPGD detectors is the use of Cu in the electrodes, which has a relatively large radiation length. The charge sensitive elements of the readout are usually made of Cu strips (or pads). Replacing these Cu strips with Al strips will yield a decrease in the detector material. This is being investigated within eRD108, specifically applied to Micromegas. Additionally, eRD108 is also investigating reducing the traditional woven Inox mesh of the Micromegas, which is used to create the electron amplification, with thin ($\lesssim 40\mu m$) layers of Al foils. If successful, this would reduce further the Micromegas material budget.
\subsubsection{Readout Structures}\label{sec:readout} 
\paragraph{{\bf{Channel Reduction}}}
The EIC will require excellent spatial resolution, traditionally achieved through finely pitched charge-sensitive strips or pads, which for large-area detectors would require lots of channels to be read out. This not only makes building, operating, and analyzing the detector more cumbersome due to its many channels, but also significantly increases its cost. There are several R\&D efforts ongoing, which began with eRD6 and are now being further developed by eRD108, to lower the required channel count while keeping excellent resolution. The goal for this R\&D is to obtain $100\;\mu m$ spatial resolution for large detectors ($\sim 5,000\; cm^2$) using a readout strip pitch of $1\; mm$.
One way to achieve this is to exploit the charge-sharing behavior of particular readout strip structures used to read out the detector. This has been demonstrated using a 1D zigzag strip structure~\cite{ZHANG2018184,9115073,9638499} to achieve a better resolution than suggested by the wide pitch of the strips via the sensitive charge sharing among the strips. Being able to have a coarser strip pitch, but achieve the resolution of straight strips with a finer pitch reduces the number of detector readout channels. R\&D activities~\cite{ZHANG2018184,9115073,9638499} testing 1D zigzag strip readouts on small MPGD prototypes ($\sim 100\;cm^2$) have already demonstrated spatial resolutions around 50-100 $\mu m$ are are achievable with up to $2\;mm$ strip pitches. Scaling up the detector size and implementing a 2D version of the zigzag strip structure are being investigated. 

Developing flexible capacitive sharing readout structures is another way that the channel count of a detector could be reduced while maintaining excellent spatial resolution. These readout structures consist of a vertical stack of alternating layers of charge-sensitive strips (or pads) and Kapton. As one moves away from the amplification region, the pitch of the strips becomes larger with each layer, with only the bottom layer (widest pitch) being read out. The various layers are capacitively coupled, so a charge in the top layer (closest to the amplification layer) will induce a charge in the layer below it, which in turn induces charge in the next layer below, and so on. The resolution of the readout is then set by the narrow strip pitch of the top layer, but the number of channels needed to be read out is set by the number of strips in the wider strip pitch of the bottom layer. There is a diamond-like carbon (DLC) resistive layer on top of the structure which serves to evacuate charges and contributes to spreading out the initial charge to aid the charge sharing among strips. Initial tests using a small scale triple-GEM prototype ($\sim 100\;cm^2$) equipped with a 2D capacitive sharing pad readout which used $0.52\;mm \times 0.52\;mm$ pads with a pitch of $6.125\; mm \times 6.125 \;mm$ structure as the top layer, and read out the bottom layer which consisted of $9.99\; mm \times 9.99\; mm$ pads with a pitch of $10\;mm \times 10\;mm$. A spatial resolution of around $250\;\mu m$ was achieved~\cite{CapKondo21}. Ongoing R\&D efforts within eRD108 are pursuing capacitive sharing strip readouts and equipping large area ($\sim 5,000\;cm^2$) detectors with them. 

%
\paragraph{{\bf{Resolving Track Multiplicity Ambiguities}}}
The EIC can also benefit from readout structure R\&D related to minimizing `ghost hits' or track multiplicity ambiguities. While the EIC is not expected to have a large track multiplicity ($\lesssim$ 10 tracks/event), being able to resolve such ambiguities will aid in the track reconstruction efficiency and performance (see Fig.~\ref{fig:xyu_proto}). One solution would be to make use of a three-coordinate readout structure to provide a correlation between three coordinates. These three coordinate structures are well suited for implementing into a capacitive sharing readout structure~\cite{CapKondo21}. Another option to investigate would be the use of fast electronics, for example ASICs with a faster sampling rate, to use the timing of the hits to remove or separate ambiguous hits.  
\subsection{Future MPGD R\&D related to EIC}\label{sec:future}
The tracking performance of MPGD detectors can be improved by further reducing their material to minimize multiple scattering and dead areas that are in the tracking acceptance. The spatial resolution of the readout is limited by the spatial resolution inherent to the MPGD detector itself, which is driven mainly by fluctuations on the position of the ionization cluster in the drift gap. Investigating the size of the drift gap and constructing detectors with several narrower drift gaps could lead to better spatial resolution.

Although excellent momentum resolution via tracking at the EIC is critical to completing its physics program, so is the PID ability of an EIC detector. Implementing tracking detectors that are also capable of performing PID would be beneficial to the EIC. Detectors such as MPGD-based transition radiation detectors (TRDs, Sec.~\ref{sec:TRD}) and time-of-flight (IF5 White Paper 1) detectors would make excellent additions to the EIC providing both hit points and PID capabilities. These types of detectors would rely on the development of fast electronics (e.g.\ sampling rate $\sim100 MHz$) at reasonable cost per channel and MPGDs with excellent timing resolutions ($\lesssim 1 ns$). Additionally, photon detectors can also be paired with MPGDs resulting in MPGD based RICH detectors (Sec.~\ref{gaseous-photon-detectors}).  
%
\subsection{Need for a Nuclear Physics MPGD Development Facility}
The MPGD tracking requirements for the EIC require large-area, low mass, and high precision space point resolution. Furthermore, the EIC would greatly benefit from PID capable MPGD detectors, which could provide large-area and low-mass detectors capable of providing high precision hit points with PID information. Implementing a dedicated facility in the US for the development, testing, and dissemination of MPGD technologies will strongly benefit the development of MPGDs needed for the EIC. Such a facility could be modeled on the Gaseous Detector Development (GDD) facility at CERN or the SiDet facility at FNAL, which is based in the US and develops silicon detectors for HEP. Although the GDD  facility at CERN serves the global MPGD community, due to their geographical proximity and strong affiliation to the CERN-based RD51 collaboration for MPGD technologies, MPGD groups involved in HEP experiments at CERN, European universities and research institutions have benefited more. A similar facility based in the US would serve as a center for the US MPGD community to coalesce around, strengthening what is already a growing community of detector experts. 
\newpage
\section{MPGD technologies for Low Energies Nuclear Physics}
\subsection{Introduction}
%
The Facility for Rare Isotope Beams (FRIB) at Michigan State University (MSU) is at the forefront of experimental low-energy nuclear physics. FRIB's mission is to provide the necessary expertise and tools from nuclear science to meet national needs, which will allow advances in our understanding of the fundamental forces and particles of nature as manifested in nuclear matter. FRIB will be the world’s most advanced facility for the production of rare isotope beams (RIBs), capable of making a majority (around 80\%) of the isotopes predicted to exist available for experiments; this opens exciting perspectives for exploring the uncharted regions in the nuclear landscape. Its scientific impact will span from the discovery of more than 3,000 new isotopes to constraining nuclear astrophysics processes, improving our understanding of open quantum systems at the limits of stability, and allow researchers to deepen the understanding of the atomic nuclei and their role in the Universe. The unprecedented discovery potential of experiments with RIBs at FRIB can be realized by implementing state-of-the-art experimental equipment that can study these isotopes at the highest rates produced.\\ 
Gaseous avalanche readout technologies, and in particular MPGD-based readouts, play an important role for the success of the science program at FRIB. A few examples of applications of MPGDs to experimental RIBs physics include low-pressure drift chambers for tracking and particle-identification (PID) at the focal plane of high rigidity spectrometers~\cite{Cortesi_2020}, position-sensitive readout for Time-Projection-Chambers (TPCs) operated in active-target mode with re-accelerated radioactive beams~\cite{bazin_low_2020}, and TPCs for the detection of exotic decay modes with stopped radioactive beams~\cite{FRIEDMAN201993}.\\ 
Challenges for the future MPGD technologies applied to the field of low-energy nuclear physics with RIBs include improved spatial resolution and segmentation; better reliability and radiation hardness while minimizing power and cost; low ion-back flow for minimized secondary effects and an increased counting rate capability; integrated electronic readout to reach high channel density, fast data processing and fast data storage.
\subsection{Requirements of low-energy nuclear physics experiments at FRIB}
%
Originally the MPGDs were developed in the framework of high-energy physics (HEP), with the aim of overcoming limitations of wire-based readouts~\cite{TITOV200725}. As soon as MPGD technologies matured and evolved in a full plethora of diversified architectures, their applications expanded to other fields of basic research and applied science, including experimental nuclear physics, astrophysics, neutrino physics, material science, neutron detection, homeland security, medical imaging and industry~\cite{Sauli_book}.\\ 
The success of the MPGD technologies lie in the capability to offer large flexibility, so that geometry and performance can be tailored to satisfy specific requirements, which plays a crucial role in the development of instrumentation for the next generation of particle colliders. However, the requirements of HEP experiments are fundamentally different from the ones that characterize the domain of low-energy nuclear physics with RIBs. The implementation of MPGDs outside the HEP from which there were conceived, requires often a compromise between detector optimization, performance and development effort (namely cost). The main differences between HEP experiments and RIB physics in the low-energy domain may include: 
\subsubsection{Rate and multiplicity}
In typical HEP experiments, the complexity of nucleon-nucleon collisions at relativistic energies and the occurrence of possible new phenomena is expected to modify the global characteristics of the collisions, so that charged-particle multiplicity is one of the important global parameters used to extract the interaction-dynamics. This requires high-precision measurements of detectors with a continuous sampling of recorded events, where up to thousands of tracks of secondary particles are generated from each interaction.\\
On the other hand, low-energy RIB physics exploits relatively simple and well understood reactions where only a few reaction products are involved in the process, including resonant elastic and inelastic scattering of protons and alpha-particles, one- or two-nucleon transfer, Coulomb excitation, fusion-evaporation reactions, and exotic radioactive decay relevant to the nuclear astrophysics science, such as beta-delayed charged particle emission. In addition, because of the low intensity of RIBs is many orders of magnitude lower than that of stable beams, together with extremely small cross sections, the typical luminosity of these experiments is relatively small. As a consequence, the efficiency of the experimental setup and the triggering strategy for the electronics become a determining factors. 
\subsubsection{Gas Gain and dynamic range}
Central tracking detectors (i.e. TPC) for charged particle momentum determination, electromagnetic/hadron calorimeters for energy measurement, muon tracking systems, Ring-Imaging Cerenkov (RICH) detectors for PID, and other HEP collider detectors are either based on the recording of signals produced by minimum ionizing particles (MIPs), which have extremely low specific ionization density, or by the recording of photon-generated single photoelectrons from liquid/solid converters (i.e. photocathode or liquid scintillators). As a consequence, the gas avalanche readout requires relatively high gas-gain to generated a measurable signals, while the detector is operated in standard gas mixtures at atmosphere pressure, more often in highly ionizing background. To achieve detection efficiency in these conditions, a suitably wide dynamic range or the suppression of the background is necessary.\\ Typical RIB experiments are based on the detection of heavy ions that release a large amount of primary ionization electrons in the sensitive volume of the detector, so that moderate gas gain is generally sufficient to achieve the required detector sensitivity. However, the reactions often involved particles characterized by very different mass and different charge, emitted within a wide range of energies, so that a wide dynamic range became a crucial requirement for the operational stability of the device and for reaching a good efficiency in extracting the kinematic properties of the reaction under study. To allow the tracking of very heavy reaction components low operation at low pressure became also an important requirement, together with low-budget material to reduce straggling and loss of resolution. 
\subsubsection{Size, complexity and versatility}
Most of the modern particle physics detector systems are prominent because of their size, complexity, capabilities and, of course, also in costs. While HEP experiments are designed to run for decades under more or less the same operational conditions, feasible RIBs experiments can accumulate enough statistics for a significant measurement within a reasonable beam time, typically from some days to a week. Generally, experiments with RIBs make use of multi-purpose gaseous tracking systems which are employed for a diversified portfolio of beams, types of reactions, wide energy range (from stopped beams to a few hundred of MeV/u), and operational conditions (form a few tens of Torr up to 1 atm) - these conditions impose a large versatility and a easy access to reconfigurability. Because of the low-energy domain and the short particle range, the detectors require smaller volumes compared to the ones used in HEP domain, with a much lower complexity even when coupled to ancillary systems.\\
Moreover, for all gaseous detectors the choice of the counting gas that fills the vessel is a delicate matter, generally selected depending on the desired mode of operation, working conditions and type of measurement. Most chambers run a mixture of noble gas (e.g. Ar) and a smaller fraction of a complex, polyatomic molecule, generally hydrocarbon (CH$_{4}$, CF$_{4}$, CO$_{2}$). The first component allows multiplication at voltage bias and is chemically stable; the second one is added to absorb photons emitted by excited atoms in the avalanche when they return to the ground state, and suppresses secondary emission allowing high gas gains before discharge. However, in AT-TPC, the fill gas is the counting gas as well as the reaction target, and it requires high purity to maximize the luminosity and and low pressure to track heavy charged products. The operation of position-sensitive gaseous avalanche multipliers in these conditions (no quencher and low pressure) is problematic because of photon-mediated secondary effects strongly limits the performance in terms of gas gain, energy resolution, long-term stability and localization capability.\\

The modality and properties of the low-energy nuclear experiments with RIBs require a new class of MPGD architecture, innovative MPGD-based readouts, and dedicated front-end electronics, specifically developed to satisfy the requirements of these experiments. 
\subsection{Present MPGDs-based detectors at FRIB}
\subsubsection{Drift chamber for tracking at the focal plane of high-rigidity spectrometers}
The S800 superconducting spectrometer is used for studying nuclear reactions induced by radioactive beams with energy between 10 and 100 MeV/u. It was in operation at the National Superconducting Cyclotron Laboratory (NSCL) since the end of the ‘90s, and it will continue to serve the nuclear physics/astrophysics community for experiments with rare isotope beams also during FRIB era. The spectrometer was designed for high-precision measurements of scattering angles within 2 msr, combined to a large solid angle (20 msr) and a large momentum acceptance (6\%)~\cite{Baz03}. The large solid angle and the high-resolution (1/10,000) were optimized for a magnetic rigidity of up to 4 Tm.\\
A crucial component for the performance of the S800 spectrometer is the focal plane detector system, which consists of an array of various detector technologies for trajectory reconstruction as well as PID. The tracking system provides a measurement of the transverse positions and angles of charged particles on an event-by-event basis, with 100\% particle transmission. It consists of two large-area Drift Chambers (DCs) placed 1 meter apart, which provide a position resolution of 0.5 mm (sigma). Under these conditions, the angle of the particle trajectory can be determined with a resolution of better than 2 mrad, even when multiple scattering effects in the tracking detector are considered.\\ 
The new upgraded detector design is based on a simple and robust gas avalanche readout scheme comprising a position-sensitive Micromegas board~\cite{GIOMATARIS:199629} coupled to a multi-layer M-THGEM~\cite{cortesi2017} used as a pre-amplification stage. This hybrid readout configuration provides a stable high-gain operation at low pressure (below 50 torr), suitable for heavy-ion tracking, by combining the unique feature of sharing the total gas gain between a cascade of elements, each one operated below the critical voltage for discharges. Signals are processed by a digital DAQ based on GET electronics~\cite{POLLACCO201881}, which provides the possibility to process multi-hit events. Fast filling gas and multi-hit capability allow for an overall increase of the counting rate capability, up to 20 kHz.\\
The successful development of the new tracking system for the S800 spectrometer will allow to carry out the rich program of nuclear physics with exotic beams at FRIB, and would also serve as a prototype for a similar, bigger tacking system needed for FRIB’s High Rigidity Spectrometer, currently in construction, and other similar spectrometers.
\subsubsection{Time-Projection-Chambers operated in Active-Target mode (AC-TPC)}
Time-Projection-Chambers operated in the Active-Target mode (AC-TPC) provide full solid-angle detection coverage, excellent energy and angular resolutions and separation of charged particle tracks~\cite{Ayyad:2018kqp,Ayyad:2020djq}. This results in a high luminosity ideal for experiments with weak beam intensities, down to a few hundred particles per second, therefore extending the reach of nuclear studies to more than a thousand exotic and rare isotopes (RIBs) that can only be produced at such low intensities. These experiments imply the use of inverse kinematics in which the the heavy reaction partner is the beam, and the target is a light particle such as proton (p), deuterium (d), 3He, or $\alpha$-particles (4He).\\
The AT-TPCs offer unique detection possibilities and qualities for the resulting low energy recoils. Hence the past decade has seen a rapid increase in the number of these types of detectors used in low energy nuclear physics experiments with RIBs. In response to the demanding challenges in terms of position-sensitive readout performance, new MPGD structures are studied and developed – such as multi-layer THGEM configurations with the mesh embedded as inner electrodes~\cite{deOlivera:2018tyv}. In addition, we are currently focusing on the implementation of a new production technology, the additive manufacturing technology for large-scale fine gas avalanche structures~\cite{Randhawa:2019cqe}. The implementation of new substrate materials for hybrid MPGD configurations are part of present developments to improve the resolution and reliability. This will reduce the discharge probability for operation of the AT-TPC readout in pure elemental gases at low pressure, as needed in the active target mode, at a reduced ion back-flow for high rate applications with highly ionizing beams RIBs in the 1-50 MeV/n energy domain. An example of recent progress is the development of the Multi-layer Thick Gaseous Electron multiplier (M-THGEM) operated now routinely as a first amplifying device combined with a Micromegas position-sensitive readout of TPC operated in active-target mode. This hybrid combination was successfully used for experiments with pure H$_{2}$, D$_{2}$ and He.
\subsubsection{The Gaseous Detector with Germanium Tagging (GADGET)} 
Beta decay may be used to populate nuclear resonances of astrophysical interest and measure their branching ratios for low-energy proton and alpha particle emission. The charged particles have commonly been detected using silicon detectors, which are unfortunately quite sensitive to energy deposition from the beta particles. Another challenge is ensuring that the low-energy particle transitions proceed to the ground state, which is usually the most relevant for astrophysics. The Gaseous Detector with Germanium Tagging (GADGET) at FRIB was designed to address these issues by depositing the beta decay parent into a gaseous TPC (which is nearly transparent to the beta particles)~\cite{POLLACCO2013102} and surrounding it with an array of high-purity germanium detectors (to detect particle-gamma coincidences from the population of excited final states ~\cite{FRIEDMAN201993}. GADGET has operated for several years in its first phase using a coarse-grained MPGD with 13 pads as a calorimeter to measure weak, low-energy (200-300 keV) proton branches, of relevance to nucleosynthesis in classical nova explosions~\cite{Friedman2020,sun}. In order to study a special case of multi-particle emission of relevance to Type I X-ray bursts on accreting neutron stars~\cite{Wrede,Glassman}, GADGET has recently been upgraded to operate in its second phase as a TPC with a higher granularity MPGD with 1024 pads and GET electronics. The GADGET TPC employs resistive Micromegas to protect the electronics from large energy depositions during the implantation of beam particles and is the first TPC in nuclear physics to be surrounded by an array of germanium detectors. TPC data analysis will be facilitated by the use of a convolutional neural network to identify events of interest and also the established ATTPCROOT analysis framework.
\subsection{Path forward and new opportunities}
A fundamental limit of gaseous detectors to both applications of TPCs and Gaseous Photomultiplier (GPM) in high-rate experiments is the accumulation of slowly-drifting, avalanche-induced ions in the active gas volume. In a TPC, an intense ion back-flow (IBF) leads to a build-up of positive ions in the drift region that causes space charge effects. This compromises the homogeneity of the electric field locally, causing a degradation of the localisation capability, and of the overall performance of the position-sensitive
readout.\\
In a GPM, the avalanche-induced ions that drift back to the photocathode (PC) cause severe limitations to the gas gain stability of the detector. Furthermore, an intense stream of positive ions that impact on the sensitive PC causes a substantial modification of the lattice defect pattern in the PC layer, resulting in a reduction of the electron escape length and a variation of the electron affinity. This eventually results in a substantial loss of quantum efficiency. \\
The gas avalanche readout is the major source of positive ions that stream back to the cathode. The amount of positive charge created during the avalanche depends on several factors, including the intensity and energy of charged particles that impinge on the detector, the type and properties of the filling gas, and the ion back-flow suppression capability of the gas amplification stage. In most traditional gaseous detector configurations, including multi-wire proportional chambers, parallel- plate counters, and resistive plate chambers, almost all avalanche ions flow back to the photocathode (GPM) or to the collection region preceding the multiplier (TPC). The IBF can be reduced by many orders of magnitude by incorporating a pulsed ion-gate electrode that takes advantage of the natural delay in the ions' arrival to block them by switching the drift field polarity. However, this comes at the expense of a considerable dead time which limits the detector rate down to several tens of Hz~\cite{Ball:2014qaa}.\\ 
The advent of advances in photo-lithography and micro-processing techniques have triggered a major transition in the field of gas detectors from wire structures to MPGD concepts, revolutionizing cell-size limitations for many gas detector applications. In cascaded micro-pattern detectors, the backflow of the positive ions is dramatically reduced, as a large fraction of the ions are collected in the intermediate elements~\cite{Shekhtman}. For instance, in Micromegas detectors a large fraction of the avalanche ions can be stopped at the micro-mesh due to the large avalanche/drift field ratio that acts as a filter~\cite{Bhattacharya:2152254}. In other detector configurations consisting of a cascade of two/three hole-type multipliers, ion back streaming can be reduced to a 1\% level by adjusting the transfer fields between successive elements. The search for new MPGD structures to further reduce the ion back-flow is one of the most pursued R\&D trends in MPGDs, and specific research activities are currently ongoing in several laboratories and research centers worldwide. Success in this field will pave the way to photon detectors with chemically-unstable photo-converters, e.g. Bialkali photocathodes with visible-light sensitivity, as well as excellent point resolution in large-volume TPCs operated at much higher luminosity. The impact of such devices will not only be tremendous for the nuclear/high energy physics community but also in applied radiation detection such as radiation monitoring and imaging.\\
One of the leading R\&D trends in novel MPGD technology focuses on embedding novel materials into present and new gas avalanche structures using new production techniques to enhance performance and stability~\cite{instruments3030051}, extremely appealing for developing applications for particle physics experimentation and other fields. For instance, Graphene is a single layer of carbon atoms arranged in a honeycomb lattice, with a reported strong asymmetry in transmission of low energetic electrons (ideally 100\%) and ions (ideally 0\%). Combining a thin graphene layer to serve as a charge filter will allow an efficient suppression of the ion back-flow (e.g. in gas photomultiplier - GPM), reduced space-charge effects in TPCs, and an efficient protection of the sensitive photocathode layers from the gas contaminants (i.e. oxygen degrades the quantum efficiency of the photocathode). Present challenges include production of uniform graphene layers, freely suspended over the avalanche volume,  as  well  as  the development of methods to measure and characterize the charge transfer properties through the graphene layer. \\
Strong collaboration with HEP groups and other fields will be crucial for fast advances in the field, and it will benefit the larger scientific community. It provides new relevant scientific and technological knowledge on MPGDs and their implementation, leading to a potentially larger application portfolio beyond nuclear physics.
\\
\\
\\
{\bf{NSLC and DOE contract acknowledgment:}} \\
This material is based upon work supported by under the National Science Foundation (NSF), under the cooperative agreement no. PHY-1102511. The AT-TPC project was partially supported by the National Science Foundation (NSF), USA under grant no. MRI-0923087. The GADGET project was supported by the U.S. National Science Foundation under Grants No. PHY-1102511, No. PHY-1565546, No. PHY-1913554, and by the U.S. Department of Energy, Office of Science, Office of Nuclear Physics under contract DE-SC0016052.
\newpage
\section{Development of Large MPGDs for High-Rate Experiment at Jefferson Lab}

\subsection{Introduction} \label{sec:intro}
The Thomas Jefferson National Accelerator Facility (TJNAF), also known as Jefferson Lab or JLab is a world-leading US national laboratory at the forefront of the study of the fundamental nature of nuclear matter. The facility is home to the Continuous Electron Beam Accelerator Facility (CEBAF), a world-class machine that deliver high intensity longitudinally polarized electron beams as a  probe for investigating nucleon structure and nuclear physics. JLab completed the 12 GeV upgrade of the CEBAF in 2017 that allows  outstanding investigation of nucleon structure in terms of form factors (FFs), transverse momentum distributions (TMDs), generalized parton distributions (GPDs) and structure functions in the valence quark region. CEBAF is able to deliver longitudinally polarized electron beams at an intensity of up to 85 $\muup$A, corresponding in some experiments to luminosities as high as 10$^{39}$ electron/s-nucleon/cm$^2$. CEBAF provide the electron beam to four experimental halls, labelled Hall A, Hall B, Hall C, and Hall D. Each hall has a set of highly specialized spectrometers to record the products of collisions between the electron beam to a fixed target to study the interaction of the quarks and gluons that make up protons and neutrons of the nucleus.
\subsection{MPGD technologies at JLab} \label{sec:mpgdAtJlab}
\subsubsection{Overview of Current MPGDs in Experiment at JLab}  \label{subsec:overview}
The challenges from the increased luminosity imposed by the 12 GeV upgrade of the CEBAF at JLab required the upgrade of drift chambers used for tracking in most spectrometers of the experimental halls with large area Micro Pattern Gaseous Detectors (MPGD) technologies~\cite{SAULI:1997nim,GIOMATARIS:199629,Bencivenni:2015} in order to cope with expected huge particle rate and to provide excellent spatial resolution ($\sim$100 $\mu$m) in order to carry out JLab's current and future ambitious nuclear physics program. The large acceptance spectrometer CLAS12 in the experimental Hall B has chosen cylindrical Micromegas detector (MVT)~\cite{mvt_jlab} technology as part of its vertex detector in combination with silicon layers (SVT). The Super Bigbite Spectrometer (SBS) experimental program~\cite{sbs_jlab}, in hall A opted for large area (200 cm $\times$ 60 cm) triple-GEM detectors~\cite{Gnanvo:2014hpa}, for the tracking in the both electron arm (Bigbite) and hadron arm (SBS) spectrometers. GEM technology offers the capability required to cope with the exceptionally high rate of a few hundreds kHz/cm$^2$ expected with SBS experiments in hall A. Large GEMs were also recently used in Hall B for tracking in the Proton Radius (PRad) Experiment~\cite{prad_jlab} and as cylindrical readout layers of the BoNUS12 TPC~\cite{bonus12_jlab} for recoil proton tracking. Future experiments at JLab heavily rely on large area and low mass MPGD technologies for tracking.
The MOLLER~\cite{moller_jlab} and SoLID~\cite{solid_jlab} programs in Hall A will use large and low mass GEM trackers. The TDIS collaboration~\cite{tdis_jlab} is developing a novel concept of modular Time Projection Chamber (mTPC) with GEM readout for recoil proton detection around the target where the background particle rate is huge. More recently, Hall B has started investigating the possibility to replace the CLAS12 forward drift chambers with large $\muup$RWELL detectors~\cite{clas12urwell_jlab} and PRad-II (an upgrade of the completed PRad experiment) will also look at the $\muup$RWELL option for its tracking layers. In Hall D, transition radiation detectors based on GEM amplification and readout (GEM-TRD)~\cite{BARBOSA2019162356} is being developed to provide additional e/$\pi$ separation capability to the DIRC and the electron calorimeter in the forward direction of the GLUeX detector. The challenges posed by the development of all these MPGD technologies for future applications at JLab are explored in more detail in the subsection~\ref{sec:challenges}.

\subsubsection{MPGDs Needs for Future Experiments at JLab: Common aspects \& differences with HEP} \label{subsec:np_hep}
MPGD technologies are widely used in high energy physics collider experiments like in the major LHC detectors ATLAS, CMS, and LHCB, typically as muon chambers or as readout planes for TPCs, whereas in nuclear physics (NP) fixed target experiments such as the one conducted at JLab, MPGDs are mostly used as tracking detectors in the  spectrometers. Because the applications of MPGDs are different for HEP and NP, the requirements on detector design, choice of material and  support structure in order to minimize material thickness while ensuring mechanical stiffness and radiation hardness for stable operation in high radiation environment are also radically different. In both HEP and NP cases,  large area, high rate capabilities, radiation hardness and space point resolution are common challenges that need to be addressed. A major difference is the emphasis on low mass (or low material budget) requirement for MPGD trackers in NP application, with radiation length typically less than  $\sim1\%\;X/X_0$ per tracking layer being of critical importance not only for minimizing background rate from mostly low energy photon conversion but more importantly to minimize multiple coulomb scattering for momentum resolution consideration. This  stringent requirement of low mass detectors is less of an issue in HEP experiments because for muon detectors, the background rate is reduced and also muons are less sensitive to material thickness, whereas for TPC detectors, the anode readout planes are located outside of the sensitive volume. The implications in term of the design and operation of the MPGD detectors and the need for separated targeted R\&D efforts for NP and HEP should  not be underestimated even though there might be strong synergy for these R\&D . As an example to illustrate the importance for targeted MPGD development for NP, the development of the NS2 technique for mechanical assembly of triple-GEM championed by the CMS collaboration for CMS Muon chambers upgrade would not be suitable for tracking in spectrometers in any of JLab experimental halls. The same argument is valid for the development of Micromegas Muon chambers for ATLAS New Small Wheel (NSW) that are not compatible with JLab environment because of the heavy support structure that these detectors required.
The development of CMS GEMs and ATLAS Micromegas required each up to 10 years of sustained R\&D effort, but are not a good fit for low mass tracking needs of JLab. It is important for the nuclear physics community and specially at JLab, to understand the need for and commit to the R\&D effort necessary to optimize MPGD technologies to the specific requirements imposed by the experiments.
\subsection{Challenges and R\&D efforts for future experiments at JLab} \label{sec:challenges}
\subsubsection{Tracking in High Rate Environment} \label{subsec:trk}
\begin{figure}[htb]
\centering
\includegraphics[width=0.85\columnwidth,trim={0pt 0mm 0pt 0mm},clip]{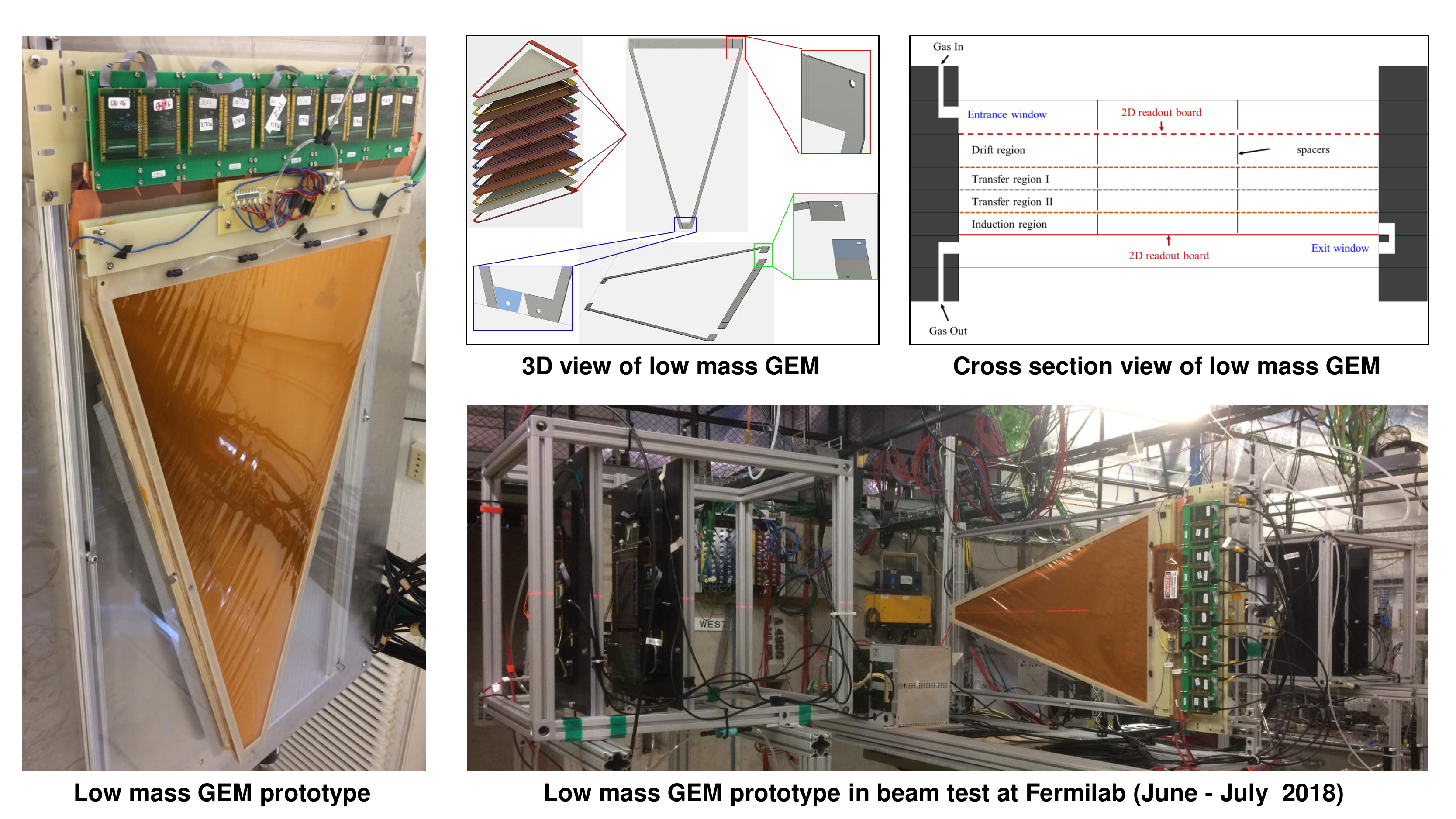}
\caption{\label{fig:eicProto} EIC "Only-foils" triple-GEM prototype (0.45\% X$_0$ / X) i.e. the sensitive area of the detector is made of only the GEM, U-V strip readout, cathode and entrance windows foils with no rigid PCB support material. The prototype is shown on the left and bottom right in beam test at FNAL (2018). On top right are the 3D sketch of the parts of the detector and the cross section view of the only-foil design.}
\end{figure}

\paragraph{\bf Development of large area \& low mass MPGD structures}
One critical consideration to satisfy momentum resolution requirements and minimize the background rate to successfully complete the JLab experimental nuclear physics program is the design of a large area tracking system with the low material thickness. This is a  design requirement not only in the active area of the detector, defining its acceptance, but also regarding the material selected for the mechanical support structure at the peripheral of these detectors. For the tracking stations in spectrometers at JLab, we typically want to develop large MPGD detector modules in order to optimize the active-to-dead area ratio in the overall acceptance of the spectrometer. This is usually in conflict with the need for strong support structure which in turn results in bulky and heavier structures.  As an illustration, a meter-size triple-GEM detector typically has 15 mm thick and 30 mm wide G10/FR4 frames at its peripheral to keep the GEM foils stretched and flat. This is in addition to 3 mm thick honeycomb sandwiched between thin G10 skins on which the GEM and strip readout foils are stacked. Though this triple-GEM design with a radiation length of $\sim 1\%\;X/X_0$ in the active area, known as the COMPASS GEM was considered as a low mass detector a decade ago,  the requirements for low mass tracking detector for new equipment such as SoLID GEMs in Hall A or future detector upgrade such as the High Luminosity CLAS12 $\muup$RWELL forward trackers are more ambitious, aiming at a factor two reduction of the radiation length (i.e. $\sim 0.4\%\;X/X_0$) per tracking layer and the width of the support frames by a factor 4  with respect to the current state of the art.  One way to achieve these requirement for triple-GEM detector, would be to build on the successful development of "Only-foils" triple-GEM concept developed as part of the EIC detector R\&D effort (eRD6) ~\cite{eRD}. The design of this detector  completely remove all rigid PCB support structures such as the honeycomb plate and the thin G10 skins from the detector active area, resulting in only GEM, cathode and strip readout foils in the detector acceptance area. A one-meter size "Only-foils" triple-GEM prototype ~\cite{lowmassGEM,lowMassGEM2} with radiation length of $\sim 0.45\%\;X/X_0$ was successfully built and operated in beam to demonstrated the feasibility of the concept.  A picture of the prototype and the beam test setup at Fermilab is shown on Fig.~\ref{fig:eicProto}. However substantial R\&D is still needed to develop new materials for light-weight strong and radiation hard support structures for triple-GEMs to operate in a stable setting in high radiation environment. We will be investigating carbon fiber structure or  narrow ceramic-based frames as alternative to fiberglass frames. At the level of the $\sim 0.5\%\;X/X_0$ radiation length, the copper electrodes of GEM foils $\muup$RWELL and of readout strips become a significant contribution of the detector's overall material thickness. A few tens of microns of Cu layers will increase significantly of the radiation length  of the detector in the active area. Similarly, the metallic mesh of a Micromegas which is based on Inox material is the major contributor to the total material thickness. The charge sensitive elements (strips or pads) of the MPGD readout structure are also usually made of Cu. Replacing the Cu in some parts of the detectors by lighter metal such as aluminum (Al) or very thin layer of chromium (Cr) will significantly reduce the detector material. Several approaches are ongoing to carry out these R\&D efforts with the investigation of 2D Al-strip readout for Micromegas detectors, and the development of Cr-capacitive-sharing readout structures for GEM and $\muup$RWELL detectors~\cite{eRD108,CapKondo21}.
\paragraph{\bf Development of high-rate resistive MPGDs}
\label{highRateResMPGDs}
The operation of MPGD structures with high gain and in spark-free or spark-protected  condition was enabled in recent years with the development of resistive MPGDs such as resistive Micromegas or $\muup$RWELL detectors. Both structures used a thin resistive layer as part of the amplification structure to quench the energy of a spark discharge and reduce the spark rate by orders of magnitude.  This feature subsequently results in a dramatic reduction of the damages to the amplification structure itself and the front end electronics (caused by sparks), as well as reducing the long dead time to recover from spark usually observed in standard MPGDs during operation. However the evacuation to the ground of the MPGD amplification charges through the resistive layer severely reduce the rate capability of resistive MPGD structures. The charge evacuation timing characteristics depends on the surface resistivity of the layer as well as size of the detector and quickly drop a few order of magnitude for large size MPGD compared to standard non resistive MPGDs. As an example a 1-meter size triple-GEM detector can easily operate in a rate environment of $\sim$1 MHz / cm$^2$, while  $\muup$RWELL or resistive Micromegas structures of a similar size with a typical surface resistivity of $\sim$10 - 20 M$\Omega \, / \, \square $, has a rate limitation on the order of $\sim$kHz / cm$^2$. A sustained R\&D effort in the future to develop large area resistive MPGD structures with high rate capabilities  ($\sim$1 MHz / cm$^2$) is therefore critically important for tracking in the high rate environments of the JLab experimental halls. Several R\&D efforts to reach high rates capability with $\muup$RWELL detectors by reducing the path for the amplification charges to the ground are been investigated by different groups all across the world. Novel high rate resistive MPGD structures should however preserve the spark resistant capabilities of such detectors and long term stability by a proper choice of the material and design. The development of large area resistive MPGD structures that combines high rate with spark resistant and radiation hardness capabilities will be extremely beneficial for the next generation of nuclear physics experiments at JLab but required a robust and sustain R\&D effort specially within by the US nuclear physics community.
\subsubsection{Development of low channel count \& high performance readout structures}
\paragraph{\bf Low channel count readout structures}
\label{lowChannelCount}
MPGD detectors typically require very finely pitched strips or pads readout plane, i.e. a large number of electronic channels to be read out in order to achieve high space point resolution capabilities because the small size of the charge avalanche cloud produced by the micro structure devices used by these detectors for signal amplification. This results not only in a significant cost of readout electronics for large area MPGD tracking detectors, but also the main driver of the event data size to be processed and transferred from the DAQ system to disks or other electronic storage during the experiment and for the offline analysis. In addition, the constraints in terms of cooling, services and connection of a large number of front-end electronics to the detector during the design, construction and operation should not be underestimated.
There are several ongoing R\&D efforts to significantly reduce the required channel count by a factor of 5, i.e. readout strip pitch typically of 1 to 2 mm for large ($\sim$ 1 m$^2$) MPGD detectors with the goal to keep spatial resolution below than $100\;\mu m$. The three main ideas currently under investigation are:
\begin{itemize}
\item {\bf Zigzag readout:} Uses interleaves of chevron-style / zigzag shape strips to facilitate charge sharing between wide adjacent strips and achieve high spatial resolution performances with low channel count~\cite{azmoun:2020tns,Zhang:2017dqw}. 
\item {\bf resistive readout:} Surface resistivity of the resistive layers of Micromegas or $\muup$RWELL detectors could also be selected to enhance the lateral spread of amplification charge cloud between large pitch neighboring strips or pads to obtain particle position with high accuracy with center of gravity reconstruction algorithm~\cite{DIXIT2004721}.
\item {\bf capacitive-sharing readout:} An alternative approach using capacitance coupling between stack of pad layers to transfer charges to large pad or strip readout. This approach presents the advantage of easy implementation in any large MPGD structures (GEMs, Micromegas and $\muup$RWELLs) and readout patterns~\cite{CapKondo21,CapKondo22}. 
\end{itemize}
\paragraph{\bf Development of 3-coordinate strip readout structures for MPGDs}
\label{threeCoord}
\begin{figure}[htb]
\centering
\includegraphics[width=0.85\columnwidth,trim={0pt 5mm 0pt 15mm},clip]{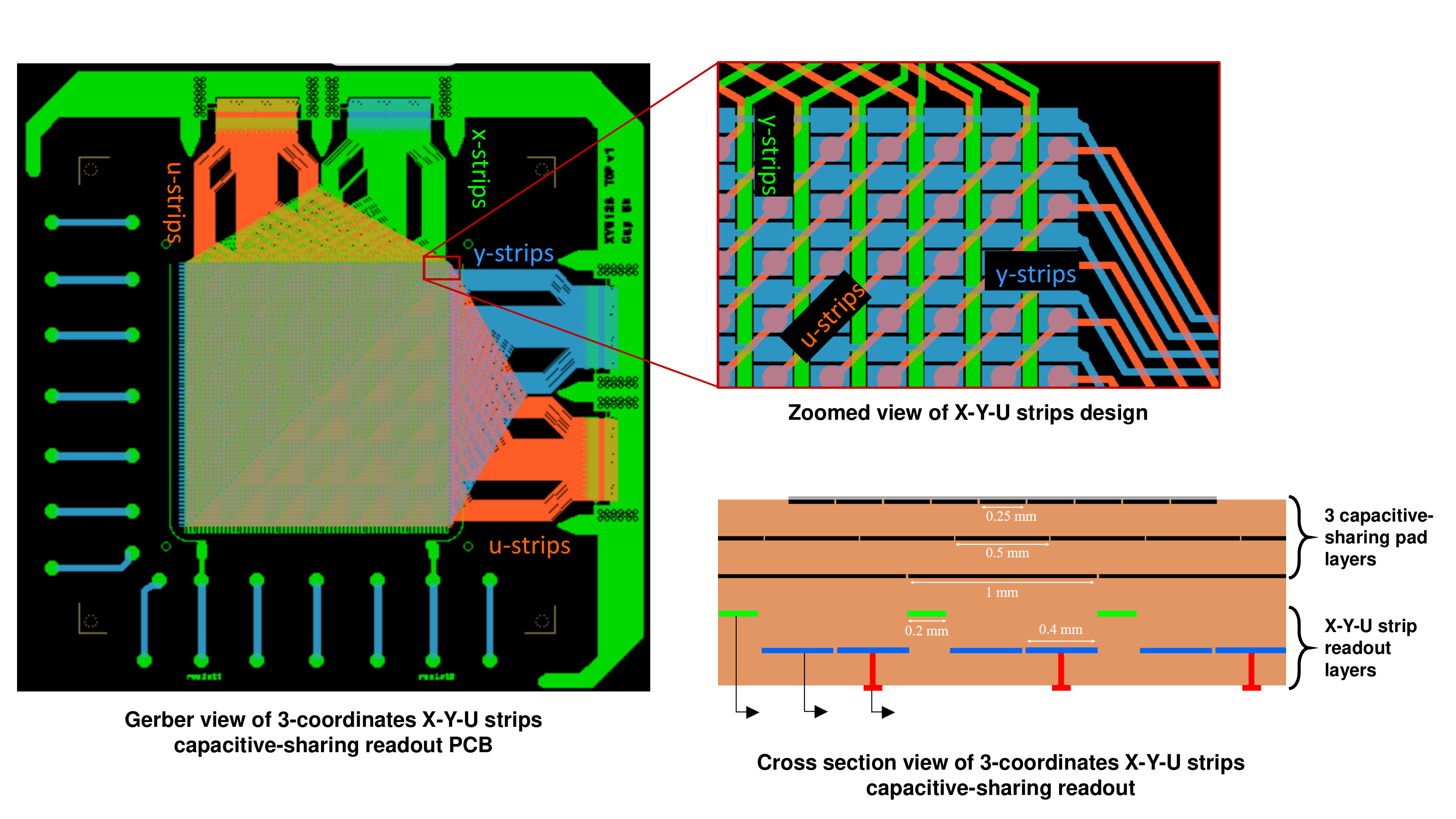}
\caption{\label{fig:xyu_proto} Design of 3-coordinates (x-y-u) strip readout PCB (\textit{left}) with a zoomed view  (\textit{top right}), showing the (x,y,u) strips configuration and the capacitive-sharing scheme shown on the cross section view on the sketch (\textit{bottom right}).}
\end{figure}
Operating large MPGD with strip readout structures in high rate environment such as at JLab will undoubtedly results in the need to address pile-up and multiple hits ambiguity ("ghost hits") issues. The problem is amplified with the development of low channel count readout structures that require large strip pitches as described in the previous section~\ref{lowChannelCount}. One elegant solution to address multiple hits ambiguity as well as improving tracking performance in high background rate environment is the development of 3-coordinates (x-y-u) strips readout structure such as the example shown on Fig.~\ref{fig:xyu_proto}. Such a strip readout structure, combined with fast readout electronics allow for the use of good timing, position and charge correlation from the signal on the set of 3 strips to accurately reconstruct the position information with high precision from a large detector operating in a high rate environment. The capacitive-sharing concept is by design the natural technology  for 3-coordinate (x-y-u) strip readout structures. A GEM prototype with (x-y-u) strip readout is under  development to demonstrate the feasibility and performances of the concept. Further R\&D will be required for the design optimization and full characterization as well as the expansion of the concept to other types of readout patterns. If successful, the combination of high-rate capability, low channel count and 3-coordinates on a single readout structure for large area detectors will represent a turning point for the development of large MPGD devices for tracking in the high background environments of JLab's experimental halls.  
\subsubsection{Development of fast timing detector with \texorpdfstring{$\mu$}{}RWELL amplification structure}
The Radiation Detector and Imaging group at JLab just recently joined PICOSEC collaboration to leverage the R\&D effort for the development fast timing detectors (tens of picoseconds) using Micromegas amplification structure~\cite{picosec}  with the $\muup$RWELL technology. We plan to investigate the use of $\muup$RWELL amplification with capacitive readout for large area (20 cm $\times$ 20 cm) PICOSEC technologies to provide good position resolution capabilities in addition to the picosecond timing capabilities. Areas that require dedicated and sustained R\&D for the optimization of large area  $\muup$RWELL-PICOSEC include the investigating option for Cerenkov radiators and stable photocathode materials as well as the development of  structures for mechanical stability and stringent uniformity requirement critical to guarantee the timing performances of the detector.
\subsection{Need for MPGDs R\&D facility in the US for the Nuclear Physics community} \label{sec:commonfac}
As highlighted throughout this white paper, the MPGD requirements for future experiments at Jefferson Lab experiments require dedicated and sustained R\&D effort to prepare for the challenges for large area, low mass, high spatial resolution and radiation hard MPGD technologies for tracking and PID in high rate environment. Having a facility in the US for the development, testing and dissemination of MPGD technologies will strongly benefit the development of MPGDs for Jefferson Lab needs. Such US-based MPGD facility could be modeled on the Gaseous Detector Development (GDD) laboratory at CERN or the SiDet facility at FNAL dedicated for silicon detector technologies for HEP in the US. Though the GDD facility at CERN is available for any MPGD groups of the particle physics community world wide, it has so far mostly benefit MPGD groups involved of the HEP experiments at CERN and in European universities and research institutions mainly because of the geographic proximity and the strong affiliation to the CERN-based RD51 collaboration for MPGD technologies. 
Jefferson Lab has in the past, vastly benefit from the knowledge, expertise and experience available within the RD51 collaboration to develop the MPGD trackers for Super BigBite program  in Hall A or BoNUS and CLAS12 in Hall B. However, with the ever increasing interest for these technologies in future experiments at the lab, a MPGD R\&D facility in the US,  coalescing MPGD communities in US national labs and universities, to address the challenges is critical for the success of the deployment of MPGD technologies in experiments at Jefferson Lab. The facility would ideally be hosted at one of DOE National Laboratory such as Jefferson Lab but made available for the entire MPGD community including NP, HEP, astrophysics as well as for medical and industrial fields.
\\
\\
\\
{\bf{DOE contract acknowledgment:}} \\
This material is based upon work supported by the U.S. Department of Energy, Office of Science, Office of Nuclear Physics under contract DE-AC05-06OR23177.
\newpage
\section{\bf MPGD Technologies for Particle Identification in Nuclear Physics Experiments} \label{sec:PID}
\subsection{The role of MPGD-based Photon Detectors in RICH Technologies}\label{gaseous-photon-detectors}
Gaseous Photon Detectors (PD) have played/are playing a major role in establishing and operating Ring Imaging CHerenkov (RICH) counters, thanks to their specific characteristics, some of them unique: they represent the most cost effective solution for what concerns the coverage of large detector areas, and they offer the minimum material budget, a feature relevant when the photon detectors have to sit in the experiment acceptance. Moreover, the gaseous PDs can operate in presence of magnetic field.
\par
The successful operation of gaseous PDs imposes to overcome two major challenges. 
\begin{itemize}
    \item 
    \textbf{Selection of the photoconverter} - The photoconverting vapours, initially used, require either extended conversion volumes (TMAE), that results in parallax errors and wide ranges of electron drift time, or very far UltaViolet (UV) detection domain (TEA). Feedback photons from the multiplication process can generate spurious hits wherever in the converting volume. They have been progressively abandoned. Among the standard solid state photoconverters commonly used in vacuum-based detectors, only CsI can be reliably used in gaseous atmosphere thanks to its relatively high work function: it can tolerate some bombardment by the ions generated in the multiplication process, where the maximum integrated bombardment before observing Quantum Efficiency (QE) degradation is of the order of 1~mC~cm$^{-2}$~\cite{CsI_ageing,CsI_ageing2}.
    \item
    \textbf{Photoelectron extraction} - In gas atmosphere, the extracted photoelectrons can be elastically back scattered by the gas molecules and be reabsorbed in the photoconverter. Effective photoelectron extraction requires specific gas atmospheres and high electric field in front of the photoconverters~\cite{photoelectron_extraction,photoelectron_extraction2,photoelectron_extraction3,photoelectron_extraction4,photoelectron_extraction5}. MultiWire Proportional Chambers (MWPC) equipped with CsI photocathodes~\cite{CsI_MWPC} have been successfully operated, for instance in HADES, COMPASS and ALICE RICHes, even if at low gain in order to limit the ion bombardment and the photon feedback. In these detectors, where the signal is due to the ion motion, low gain results in slow operation.
\end{itemize}
\par
MPGD technologies offer natural answers to ion back flow and photon feedback suppression and much faster operation, as tested by successful applications: 
\begin{itemize}
    \item
    the PHENIX HBD with triple GEM PDs~\cite{HBD};
    \item
    the COMPASS RICH upgrade with Hybrid (THGEMS and MICROMEGAS) PDs~\cite{compass_hybrid};
    \item
    the windowless RICH prototype and test beam with quintuple GEM PDs~\cite{windowless};
    \item
    the TPC-Cherenkov (TPCC) tracker prototype with quadruple GEM PDs~\cite{TPCC}.
\end{itemize}
In multiple layer GEM PDs, where the top layer is coated with a CsI film and acts as photocathodes, the photon feedback is stopped by the limited optical transparency of the stack of GEMs, while the ion backflow is reduced because part of the ions are trapped in the intermediate detector layers. In a hybrid detector including two THGEMs and a MICROMEGAS, the first THGEM is the photocathode substrate and the feedback photon are stopped as in the GEM detector. The intrinsic ion blocking characteristics of the MICROMEGAS makes possible photon feedback rates at a few percent level. These detectors can be used in focusing and proximity focusing RICHes.
\par
The major element of interest for future applications of MPGD-based PDs is in developing the concept of compact RICH for PID of high momentum particles, that would empower the application at colliders with hermetic coverage detectors and, therefore, is a must at the EIC. In fact, RICHes for PID at high momenta require gaseous radiators, as only small-value refractive index give access to PID at high momenta.  The radiator must be long to ensure the required Cherenkov photon yields. Higher photon rates can be obtained in the far UV domain, around 120-140~nm. Access to this wavelength range can be obtained by a windowless RICH~\cite{windowless}, where the radiator gas is also the PD gas. This poses specific requirements to MPGD-based PDs. FluoroCarbons (FC) are mainly used as radiator gasses thanks to their high density, that ensures good Cherenkov photon rates, and their low chromaticity, that make possible fine resolution, a need for PID at very high momenta. The FC Global Warming Potential (GWP) is extremely high and, therefore, their use is subject to increasing restrictions, also effecting procurement possibilities. A proposed alternative is by pressurized (at a few bar) noble gasses able to mimic FC in terms of density and chromaticity. MPGDs operation in FC atmosphere has been proven~\cite{windowless}, while their ability to operate in high pressure noble gasses has to be established, in spite of some positive hints from literature\cite{high_pressure,high_pressure2}. Another need for the compact RICH concept is the fine pixelization, required to preserve the fine resolution with shorter lever arm imposed by the compactness requirements. 
\par
The possibility to identify novel solid-state photoconverters providing higher QE and adequate for operation in gaseous PDs has to be pursued: it is beneficial for the compact RICH concept and, more in general, for all the applications of gaseous PDs. Hydrogenated nanodiamond powders have been proposed~\cite{ND_principle} and initial studies are ongoing~\cite{ND_development,ND_development2,ND_development3}, while further investigation is needed, with dedicated attention to novel C-materials.
\par
In conclusion, MPGD-based PDs are an option for further developments in the Cherenkov imaging techniques and, in particular, for the needed concept of compact RICHes, essential at the EIC and that, more in general,  can open the way to a wider use of RICHes in collider environments.

\subsection{MPGD-based Transition Radiation Detector}\label{sec:TRD}
\subsubsection{Physics motivation}
Electron identification plays a very important role for physics emerging at the Electron-Ion Collider (EIC). The following processes are regarded as essential for EIC physics program and could be accessed with a help of improved electron identification. Events with electrons in the final state are important signatures for DIS physics at EIC. Secondary electrons could be emitted from leptonic and semi-leptonic decays of hadrons, for such processes as J/$\psi$ production (branching ratio to $e^+e^-$ pair is  the order of ~6$\%$), D-mesons production ( with its  Br($D^+ \rightarrow e +X$) $\sim 16 \%$), and B-mesons production ( lepton  Br($B^{\pm} \rightarrow e  + \nu +X_c$ ) $\sim 10\%$)~\cite{PDG}. Electron identification plays an important role for many other physics topics, such as spectroscopy, beyond the standard model physics, etc. Isolated electrons are not easy to identify at the EIC  because of the relatively large QCD background from hadrons.
 
 The next generation of high intensity accelerators and the demand for precision measurements from the physics requires the development of highly granular detectors. A high granularity tracker combined with a transition radiation option for particle identification could provide additional information necessary for electron identification or hadron suppression. 

\subsubsection{Current experience / experiments}
Transition radiation (TR) is produced by charged particles when they cross the boundary between two media with different dielectric constants~\cite{TR-GINZB}. 
The probability to emit one photon per boundary crossing is of order $\alpha \sim 1/137$. To increase the transition radiation yield, multi-layer dielectric radiators are used, typically several hundred mylar foils, polyethylene foam, or fibers (fleece)~\cite{TRD_BD93}.
The energies of transition radiation photons emitted by relativistic particles are in the X-ray region with a detectable energy range of 3-50~keV~\cite{TR-GARIB}. These photons are extremely forward peaked (within an angle of $1/\gamma$). The total transition radiation energy emitted is proportional to the $\gamma$-factor of the charged particle. 

Typically, in particle physics, transition radiation detectors (TRDs) are used for electron identification and for electron/hadron separation. Transition radiation detectors based on multi-wire chambers or straw tubes were widely used in high energy experiments, such as ATLAS, ALICE, ZEUS, and HERMES.  

The initial concept and first design of a GEM-based Transition Radiation detector was proposed for the EIC detector R$\&$D program~\cite{eRD}, which is supported by the DOE Office of Nuclear Physics. 

It combines a Gas Electron Multiplier (GEM) tracker with the TRD functionality. A standard GEM tracker~\cite{SAULI:1997nim} with high granularity (400~$\mu$m strip pitch) capable of providing high resolution tracking was converted into a transition radiation detector and tracker (GEM-TRD/T). This was achieved by making several modifications to the standard GEM tracker. First, since heavy gases are required for efficient absorption of X-rays, the operational gas mixture has been changed from an Argon based mixture to a Xenon based mixture. Secondly, the drift region also needed to be increased from $\sim$3~mm to 21~mm in order to detect more energetic TR photons. Then to produce the TR photons, a TR radiator was installed in front of the GEM entrance window. Finally, the standard GEM readout (based on the APV25~\cite{APV25}) was replaced with one based on the relatively faster, JLAB developed, flash ADC (FADC)~\cite{FADC125}.
\begin{figure}[h!]
\centering 
  \begin{minipage}{0.45\textwidth}
    \includegraphics[width=1.\textwidth,keepaspectratio]{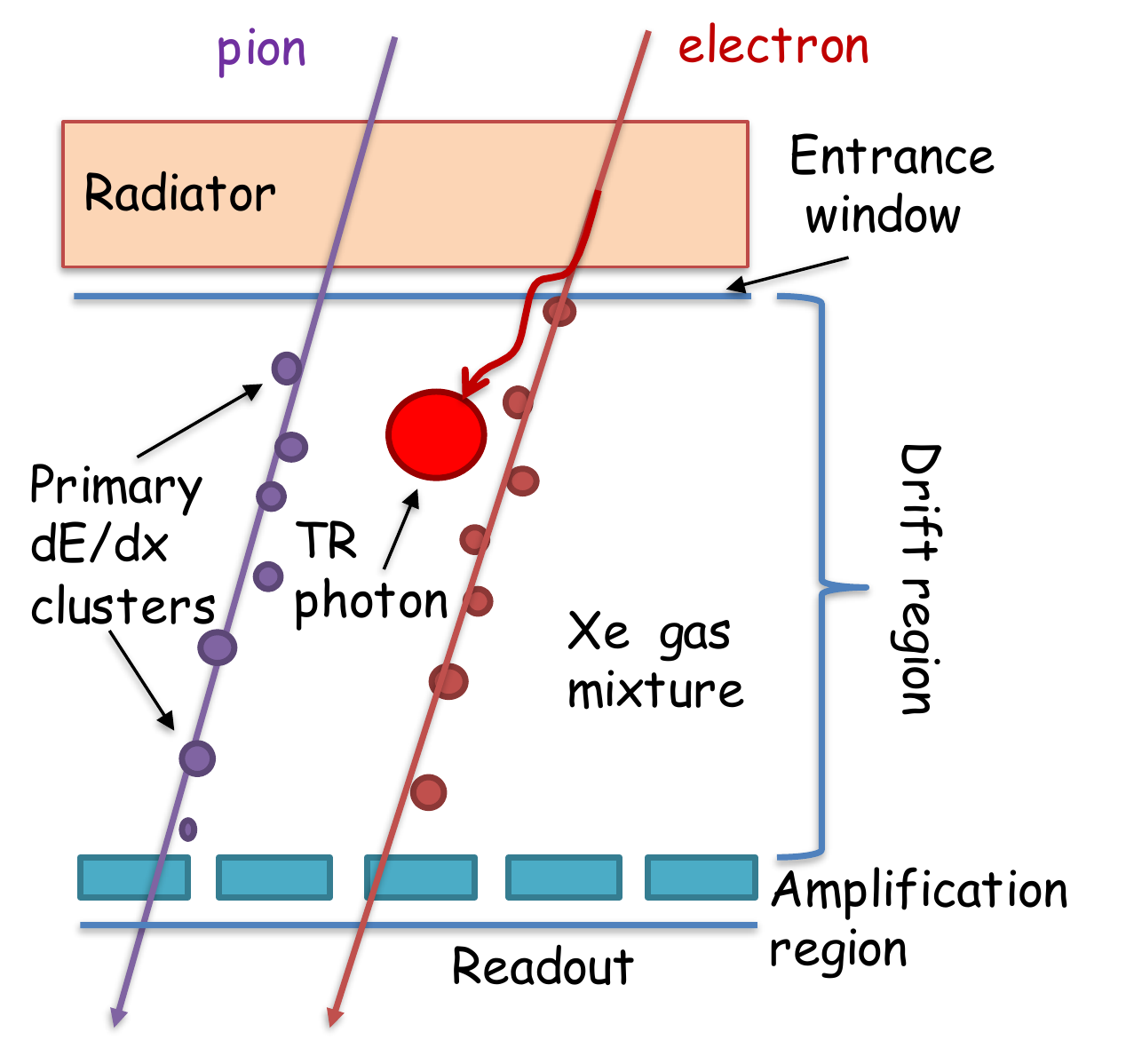}
    \caption{GEM-TRD/T operation principle}
    \label{fig:gem-con}
  \end{minipage}
  \hspace{0.1cm}
  \begin{minipage}{0.45\textwidth}
    \includegraphics[width = 1.\textwidth,keepaspectratio]{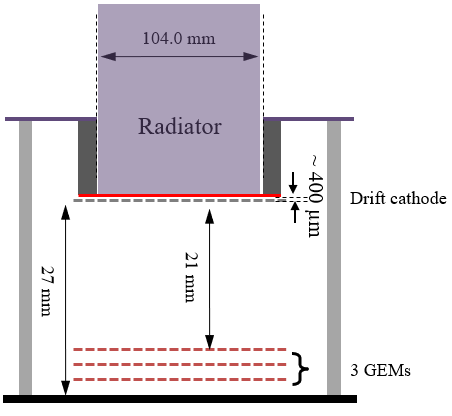}
    \caption{Schematic of GEM-TRD/T prototype}
    \label{fig:trd_draw}
  \end{minipage} 
\end{figure}
The GEM-TRD/T concept is shown in Fig.~\ref{fig:gem-con}. During the generic EIC R\&D program a small 10 $\times$ 10 cm$^2$ prototype, as shown in the Fig.~\ref{fig:trd_draw}, has been build and tested~\cite{BARBOSA2019162356}. The first beam test measurements using the GEM-TRD/T prototype were performed at Jefferson Lab (CEBAF, Hall-D) using 3-6~GeV electrons. First results show that an $e/\pi$ rejection factor of 9 can be achieved with a single GEM-TRD/T module and can be increased up to a factor 16 by using a thicker radiator (ca. $25$~cm) with 90\% of electron efficiency. 
\subsubsection{R\&D for future experiments ( prototype,readout, radiator,gas)}
We would like to continue this effort and are happy to invite experimentalist to develop MPGD-types of TRDs for the future experiments as EIC and beyond. For the successful operation of MPGD-based TRDs the following challenges need to be addressed: 
\begin{itemize}
    \item {\bf Large scale prototype.}
    During the generic EIC R\&D activity our group used a small-size (10$\times$10 cm$^2$) detector to validate the GEM-based TRD concept. However, the EIC project will require large area detectors, requiring one to perform tests on large-size MPGD based TRD modules in order be able to workout possible issues, such as noise, gain-uniformity, drift-time issues, HV stability, etc. This would also allow one to work-out design issues associated with the gas/field cage to make sure that they are suitable for TR applications.  

In order to keep the electric field uniform a special field cage needs to be developed. This includes the mechanical design and construction of the field-/gas-cage to minimize a Xe-filled gas gap between the radiator and the drift cathode. The GEM-TRD will need two HV lines: one for the GEM amplification stage and the second to set a uniform drift field. To work in a high occupancy environment, the drift time needs to be minimized, requiring fields of $\sim$2-3 kV/cm. For a 2 cm drift distance the HV should be at the level of 4-5 kV. Depending on the chosen grounding scheme, the total voltage including the GEM stage, could be up to 8-9 kV. Optimization of HV for large drift distances for a large scale prototypes need to be performed. 

Different anode readout PCB layers could be evaluated, which include capacitive-sharing pad readout and large size 2D zigzag strip readout options, and demonstrate that the concepts work equally well in a TRD application.

A low mass radiator available for mass production is critical and various materials still need to be tested and optimized. This includes the optimization of a pseudo-regular radiator using thin ($\sim12-15\,\mu$m) Kapton foils, thin net spacers, and a detailed study of available fleece/foam materials for TR-yield. 
\\
Over the past few years, the price of Xe has gone up significantly. Design and development of a re-circulation system to purify, distribute, circulate, and recover the gas, possibly based on a design of ATLAS TRD gas system at CERN, will be necessary, and should require only moderate R\&D.\\
\begin{figure}[h!]
\centering 
  \begin{minipage}{0.43\textwidth}
    \includegraphics[width = 1.\textwidth,keepaspectratio]{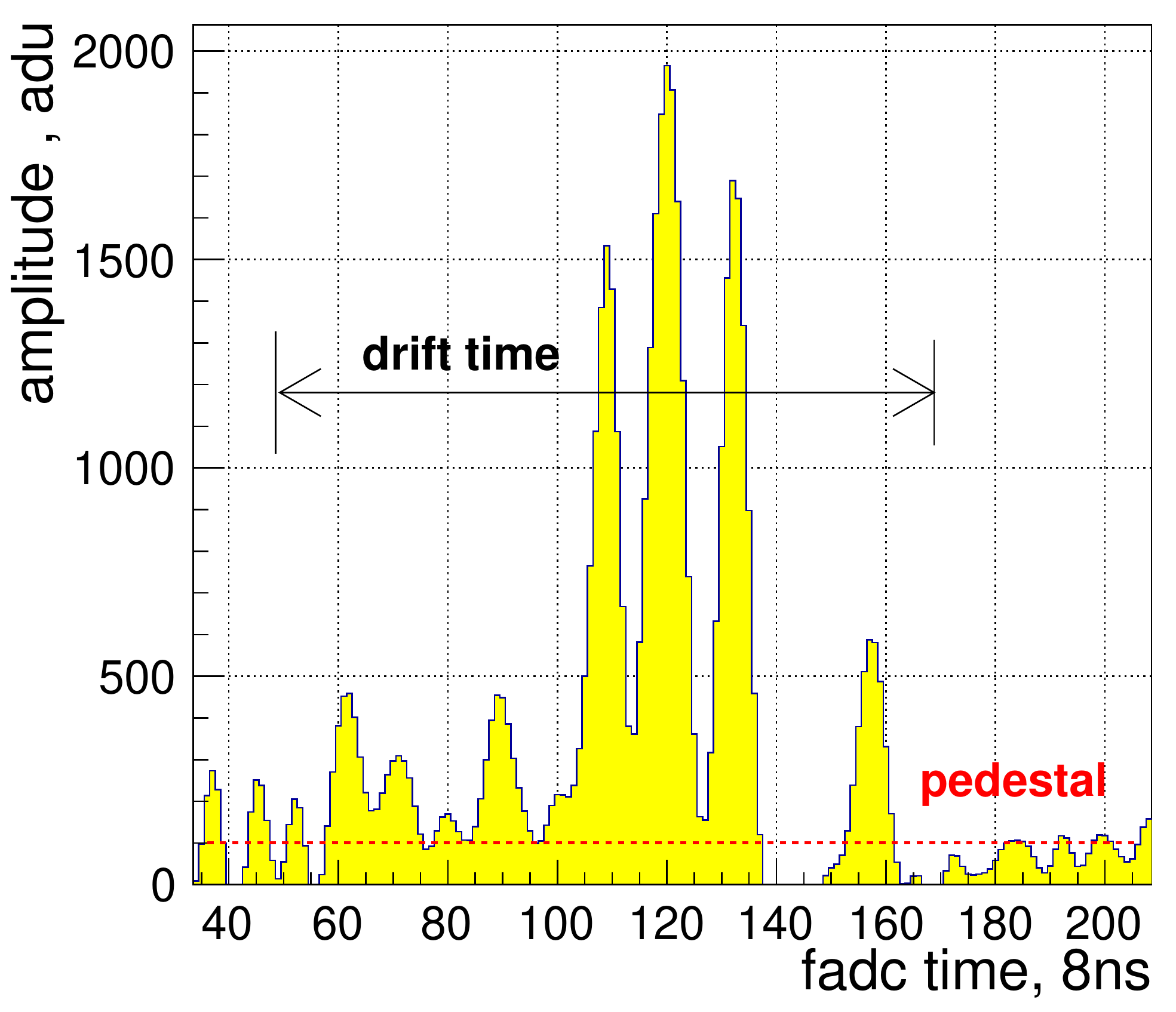}
  \end{minipage}
  \hspace{0.1cm}
  \begin{minipage}{0.43\textwidth}
    \includegraphics[width = 1.\textwidth,keepaspectratio]{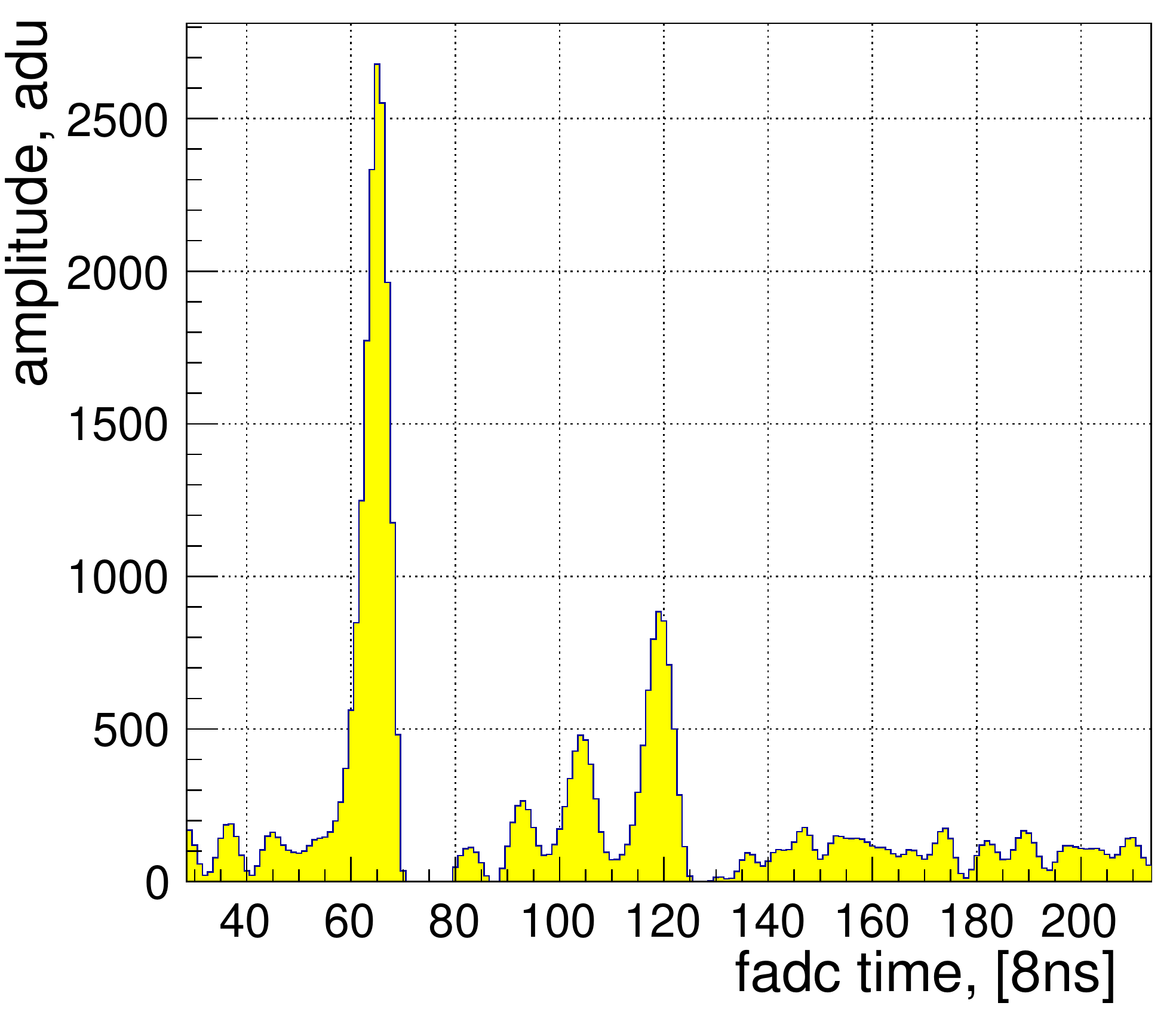}
  \end{minipage}
    \caption{ Typical flash ADC waveform }
    \label{fig:wave_fadc}
\end{figure}
\item  {\bf Readout electronics}
The standard readout for GEM detectors is typically based on the APV25 chip and measures the peak amplitude~\cite{APV25}. A TRD needs additional information about the ionization along the track to discriminate TR photons from the ionization of the charged particle. During the generic EIC R\&D tests of the GEM-TRD/T prototype we used a precise (125~MHz, 12~bit) FADC, developed at JLAB, with a VME-based readout. Those FADCs have a readout window (pipeline) of up to 8~$\mu$s, which covers the entire drift time of the GEM-TRD/T prototype. GAS-II pre-amplifier ASIC chips~\cite{FADC125} have been used, which provide 2.6~mV/fC amplification with a peaking time of 10~ns. A typical waveform signal, analyzed with the FADC system is shown in Fig.~\ref{fig:wave_fadc}. The flash ADC has a sampling rate of 125 MHz and 12 bit resolution, but provides only pipe-lined triggered readout with a total price of about \$50 per channel. The collected high resolution data recorded in test beams allow us to estimate the minimum needed shaping times of preamplifier, the FADC sampling rate and corresponding resolution. Development of a new streaming readout, similar to the one described above, will be needed to enable the streaming of zero-suppressed data over fiber links. The currently employed readout electronics will be used to formulate a final set of specifications in driving the design of an ASIC and readout in conformance with the EIC streaming readout architecture.

\item {\bf Data analysis and machine learning for e/hadron separation } To determine the electron identification efficiency and pion rejection power we tested several methods: total energy deposition, cluster counting, and a comparison of the ionization distribution along a path using maximum likelihood and neural network (NN) algorithms. The maximum likelihood and NN algorithms demonstrated similar performances. However, the NN algorithm has an advantage in practical application as it allows for the optimization of various test parameters and was used as the main analysis method when analyzing the data from the GEM-TRD/T prototype. The ionization along the track was used as input to a neural network program (JETNET~\cite{JETNET}, ROOT-based TMVA~\cite{TMVA}). Further development of Machine Learning algorithms for e/$\pi$ separation is ongoing.  
Our plan is to develop and build a functional demonstrator for FPGA Machine Learning application. A FPGA-based Neural Network application would offer real-time, low latency (~1-4 $\mu$s), particle identification. It would also allow for data reduction based on physical quantities during the early stages of data processing. This will allow us to control data traffic and offers the possibility of including detectors with PID information for online high-level trigger decisions, or online physics event reconstruction. Preliminary tests have been preformed during the generic EIC R\& D program. 
\end{itemize}
%

\newpage
\section{Electronics, DAQ \& readout system for MPGD technologies}
\subsection{Streaming readout at the EIC}

Streaming readout has been established for data processing, collection and analysis at the EIC. The consensus from the EIC user community on streaming, i.e. triggerless, readout was summarized within the Yellow Report~\cite{EICYR2021}.

\begin{figure}[h]
\resizebox{0.95\textwidth}{!}{\includegraphics{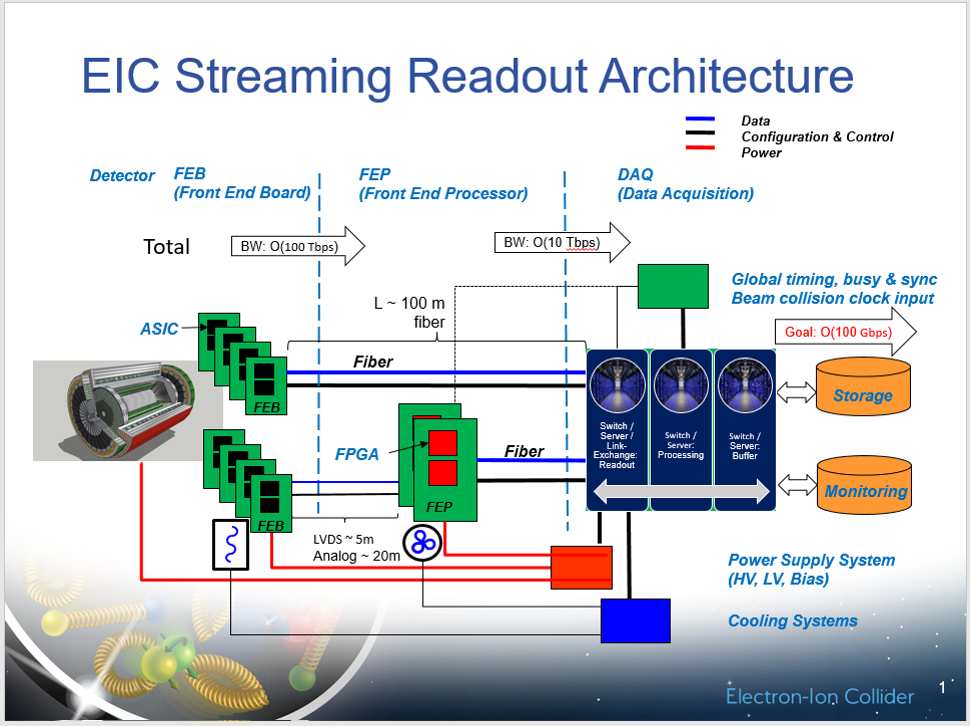}}
\caption{ The EIC Readout Architecture}
\label{fig:EICReadout2}
\end{figure}

The EIC readout architecture is shown in Fig.~\ref{fig:EICReadout2}, where the readout chain is partitioned into three distinct functional profiles: FEB, FEP and DAQ.

The FEBs, Front-End Boards, are specifically designed for, mounted on or in close proximity to each of the sub-detectors. These boards conform to the geometry requirements of each of the sub-detectors and may be further constrained by power dissipation or heat loads, cooling services, radiation levels, cabling requirements and serviceability. The FEBs are characterized by the use of low noise, low power circuitry with analog front-ends and digitization, processing and drive capabilities. This high density mixed-mode circuitry is normally implemented with Application Specific Integrated Circuits (ASIC) when a very large number of readout channels justifies development efforts; alternatively, discrete implementations with commercial-off-the-shelf (COTS) components may be developed, if permitted by the established design boundaries.

The FEPs, Front-End Processors, are located outside of the proper sub-detector volume, or in close proximity, and interface to FEBs. The FEPs aggregate the data output streams from multiple FEBs and are designed to handle different types of FEBs, i.e., a few FEP designs with generic interfaces will handle multiple sub-detectors. Some sub-detectors may not necessitate the use of FEPs, such as MAPS, as these may interface directly to switches, servers or link exchange modules (e.g., FELIX). The FEPs make extensive use of FPGAs for data processing, providing an opportunity to decrease the available output data bandwidth requirements by a factor of ten (10). This can be accomplished by various methods or algorithms, such as zero suppression or via ML and AI filtering. High performance FPGAs, available as COTS, provide the processing power, speed and flexibility of use. The choice of FPGAs and their location within the experimental area must be carefully considered with regards to radiation levels. As with the design considerations for the FEBs, the FEPs will play a critical role in configuration, control and timing of the various subsections of the readout chain via optical fibers. 

The DAQ segment will consist of Front-End Link eXchange (FELIX) boards, servers and network switches, which will be located away from the detector, facilitated by the extensive use of optical fibers. Servers and switches are available as COTS. The FELIX board, originally developed for the ATLAS experiment, interfaces the DAQ and the detector front-end and functions as a router between custom serial links and a commodity switching network using standard technologies. FELIX is designed to be detector independent, supports the CERN GBT protocol to connect to front-end units, supports distribution of timing, trigger and control and supports calibration operations. It is expected that an updated design will be necessitated to fully benefit the EIC streaming readout. Optical fibers from the FEB and FEP are employed for data transport.

With the implementation of a fully streaming readout model, it is important to provide enough available bandwidth at the FEB to handle potentially highly variable rates (signal/background/noise) from all detector systems. The total anticipated bandwidth limits for the whole detector are shown to be on the order of 100 Tbps at the front-end. With the implementation of the FEP stage we assume limits can be decreased by a factor of ten to 10 Tbps at the server/switch/link exchange level. Given current estimates on the physics interaction rates, following back-end processing and buffering, the expected data output rates to more permanent storage is anticipated to be around 100 Gbps. It is anticipated that the collision rate at the EIC will be four orders of magnitude lower than that at LHC: filtering out potential high background by means of ML/AI algorithms will be critical, however.

Timing, generally consists of global timing, busy, synchronization and clock distribution within the experimental area. Precision timing will require precision timing referencing the beam or beam crossings, as well as multiple control loops addressing stability and drift. A single accelerator clock reference may exhibit timing jitter of a few hundred femtoseconds, implemented with single-mode fibers and periodic phase correction. 

Clock distribution jitter via multi-mode optical fibers, clock de-skewing and distribution via backplanes and employing COTS components can result in clock jitter of about 4 ps, which is sufficient for detectors requiring timing resolution in the 100 ps range. As some EIC timing detectors are specified for 20 ps timing resolution, it is expected that clock jitter will need to be better, within the 1 ps range.

Consideration for calibration, testing and timing of the various subsections and components of the EIC streaming readout requires that every part of the readout chain must be designed for the intended purpose of streaming readout and also for a triggered implementation. That is, the readout system should be able to operate with and without a trigger.




\subsection{MPGD front-end electronics adapted to triggerless mode}

\subsubsection{Presently existing front-end ASICs}

Several front-end chips are presently existing to read MPGD detectors. Most of them are based on a charge-sensitive amplifier (CSA) associated to an amplifier/shaper. Signals are then either stored on a capacitor array or digitized and the data are stored in a digital buffer. The data or signals are then transferred to next element of the readout chain.

These chips are in general foreseen to be used in triggered mode, meaning that a trigger signal is delivered to the chip to trigger the readout of one event. This signal can be generated internally, based for instance on the amplified analog signals discrimination, or is transmitted from an external trigger system. In a streaming readout DAQ environment however no trigger signal is used and the front-end chip is meant to transmit to the DAQ system all signals which are read. Such a feature has to be foreseen in the chip architecture, and the output bandwidth of the chip has also to be large enough to transmit the foreseen data flux.

Only a few front-end ASICs compatible with a streaming readout DAQ are presently available. Among them the SAMPA chip and the VMM chip begin to be used in several particle physics experiment.


\paragraph{SAMPA chip}

The SAMPA chip \cite{SAMPA_chip} is a 32-channel ASIC with on-board pre-amplification, pulse shaping, digitization and DSP subsections, including a high-bandwidth digital interface for computer readout (11 links at 320~MB/s). It was developed by a group lead by the Sao Paulo University.

The chip is fabricated with TSMC 130~nm CMOS technology with a chip area of 9.6 x 9.0~mm². A charge-sensitive amplifier amplifies the measured analog signals, followed by a near-Gaussian pulse shaper. The 10-bit Successive Approximation ADC digitizes the amplified and shaped signals at a sampling rate of 10~MS/s (which can be configured to up to 20~MS/s), whereas the on-board DSP circuitry filters and carries out signal processing and compression operations on the digitized data. The chip offers a sufficiently high gain of 20-30~mV/fC with a low-noise performance (less than 1000~e$^-$). The chip is well adapted to the readout of MPGD, in particular in the TPC like the ALICE TPC at CERN, and in the front trackers.

\paragraph{VMM chip}

The VMM chip \cite{VMM_chip} was developed at Brookhaven National Laboratory (BNL) as a 64-channel mixed signal ASIC based on the IBM 130~nm technology for tracker readout, in particular the Micromegas and sTGC detectors of the ATLAS Muon Spectrometer's New Small Wheel upgrade. The chip features a novel third-order filter and shaper architecture, which results into a higher dynamic range, enabling the measurement to achieve a high resolution at moderate input capacitance ($<$200~pF), while it is also able to handle large capacitance up to 3~nF. The architecture offers a variable gain in eight values from 0.5 to 16~mV/fC with four possible shaping time intervals between 25 and 200~ns.

An excellent feature of the chip is to have both time and amplitude (peak) detection circuitry on-board. For each channel the amplified and shaped signal is passed over to both a peak detector and time detector working in tandem and giving their respective output to a digitization subsection. Output from the peak detector is given to both a 6-bit ADC for a dedicated low-delay output (50~ns delay), to be used for trigger or lower precision measurements, and to a 10-bit ADC for precision read-out, whereas the time detector has its output passed over to an 8-bit ADC for TDC functionality. Output data flux can reach a bandwidth of 1~Gbit/s.

\subsubsection{Project for a new chip in 65~nm technology}

A new initiative was recently launched by the University of Sao Paulo (Brazil) and the IRFU institute of the CEA Saclay (France) to develop a new front-end ASIC dedicated to almost all kinds of MPGD and beyond, and compatible with the requirements of modern streaming readout DAQ architectures, like the ones foreseen for the EIC project. The new ASIC, named SALSA chip, is meant to be versatile enough to cover most MPGD applications with different requirements, like large capacitance electrodes of large area track detectors, long time gates of time projection chambers, or low amplitude signals of photon detectors. Thus the ASIC shall be compatible with large ranges of signal amplitudes, electrode capacitance and propose a large range of peaking times, with optimized data processing. In order to propose such a versatile chip with a limited die size and a low power consumption, it will be designed in a more modern TSMC 65~nm technology, compared to the 130~nm technologies of the previous generation of chips. 

\paragraph{Preliminary specifications}

A preliminary set of specifications for SALSA chip was determined based on the requirements of the MPGD detectors foreseen in EIC project and in other experiments. The ASIC is foreseen to provide the readout of 64 channels adapted to a large range of electrode capacitance and with a large range of gains and peaking times (Table~\ref{table:New_SALSA_chip_specifications}), accepting both polarities of input signals. The front-end part of the channel integrates an optimized charge sensitive amplifier with tunable gain and anti-saturation circuit, a quasi-Gaussian shaper with different selectable peaking times and an adapted pole zero cancellation stage. The output of the shaper is sampled and digitized thanks to an in-channel 50~MS/s 12~bits ADC.

\begin{table}[h!]
\resizebox{1\columnwidth}{!}{%
 \begin{tabular}{||c|c||} 
 \hline
 Parameter & Value   \\ [0.5ex]
 \hline\hline
 \multicolumn{2}{||c||}{Analog characteristics} \\ [0.5ex]
 \hline
  Number of channels      & 64 \\ 
 Peaking time range      & 50 to 500~ns  \\
 Input dynamic range     & 0-50~fC to 0-5~pC  \\
 Input capacitance range & Optimized for 200~pF, reasonable gain up to 1~nF  \\
 Input rates             & 25~kHz/channel, with faster CSA reset for larger rates  \\
 Additional feature     & Reversible polarity  \\
 \hline
 \multicolumn{2}{||c||}{Digital characteristics} \\ [0.5ex]
 \hline
 ADC sampling rate      & 10 to 50~MS/s  \\
 ADC dynamics           & 12~bits  \\
 Data processing        &  Pedestal subtraction, common mode correction, zero suppression, peak finding, software trigger generation  \\ [1ex]
 Output data links      & One or a few gigabit links \\
 \hline
\end{tabular}
}
\caption{Preliminary specifications of the SALSA chip}
\label{table:New_SALSA_chip_specifications}
\end{table}

An integrated DSP will be able to process the data in order to perform basic treatments like pedestal subtraction, common mode noise correction and zero suppression. More sophisticated processing could be also done like trigger signal generation, peak finding, hit counting, or other kind of digital data filtering. Continuous readout mode compatible with a streaming DAQ will be naturally proposed, but the ASIC will be also compatible with more classic triggered DAQ. One or several gigabit output links will be available for the data transmission to the DAQ. The ASIC shall be able to stand input rates of at least 25~kHit/s/channel and beyond.

The SALSA chip die shall be small enough, around 1~cm$^2$, to be able to be installed close to the detectors, with a low power consumption at the level of 15~mW/channel. Thanks to the use of 65~nm technology and hardening techniques (TMR) the SAMPA chip will be able to stand quite harsh environment in terms of particle radiation (TID and SEUs), magnetic field or temperature.

\paragraph{Status and timeline}

The SALSA chip project was recently launched and the ASIC specifications will be soon fully finalized. Studies on the front-end and ADC architectures with the 65~nm technology are ongoing at Sao Paulo and Saclay. Some developments done at CERN on specific IP blocks for high-speed data links or internal PLL-generated clocks could be of interest in the ASIC architecture. Contacts are taken with CERN group about this topic. Development phase is foreseen to last around two to two and half years, followed by a preserial production phase of one year. Production will begin after this phase, in accordance with the EIC timeline.

\subsubsection{Needs for specific MPGD applications}

Some particular applications of gaseous detectors require readout electronics with specific characteristics which may be barely provided by readout ASICs described above. This is particularly true for the following MPGD applications.

\paragraph{Photon detectors}

Gaseous photon detectors are foreseen to be used for instance in RICH detectors, in particular in EIC experiments. Cerenkov photons are converted in electrons in photosensitive material, like Cesium Iodide (CsI), deposited on specific glass windows. The low conversion yield, at the level of 10\% induces very low signal amplitudes, close to the noise level. Readout electronics noise must be low enough in order to limit the degradation of the signal. ASICs as listed above could have a low enough noise level compatible with the requirements of the photon detectors. But for more stringent specifications very low-noise amplification-only ASICs, like some evolution of the IDeF-X chip \cite{IDeF-X_chip}, could provide a solution, with noise levels at the order of 100~electrons. However these chips would have limited functionalities and need to be associated to other chips which will take in charge the remaining of the signal treatment chain.

\paragraph{ps-level time resolution detectors}

MPGD detectors can achieve excellent time resolutions, at the level of a few tens of ps, when they are associated with a Cerenkov radiator and a layer of photoemissive material, like in the PICOSEC project \cite{picosec}. Present prototypes are using specific very fast discrete readout electronics, associated with ps-level time digitizers based for instance on the SAMPIC TDC chip \cite{SAMPIC_chip}. However these present electronics are not well integrated and require a quite large space to be installed. Projects to develop more integrated fast amplifiers and TDC are considered but this has not yet been achieved. An other solution would be to use present or in-development ASICs, but this could lead to a degradation of the time resolution of the detector as it would be limited by the sampling rates or the time resolution provided by these chips.
\newpage
\bibliographystyle{unsrt}  
\bibliography{biblio.bib}
\end{document}